\documentclass[preprint]{revtex4-1}
\providecommand{\doctitle}{~} % These are used as default values
\providecommand{\docsubtitle}{~}
\providecommand{\docauthor}{~}
\providecommand{\docversion}{~}

\usepackage{graphicx,amsmath,amsfonts,amssymb,color,
dcolumn,siunitx,bm,paralist,setspace,rotating,array,fix-cm,
multirow,hhline,enumitem,cancel,tikz,pgfplots,listings,ifthen,
framed,fancybox,empheq,verbatim}
\usepackage[utf8]{inputenc}

\empheqset{box=\fbox}

\usetikzlibrary{calc,patterns,shapes} 
\pgfplotsset{compat=newest}
\usetikzlibrary{plotmarks}
\usetikzlibrary{arrows.meta}
\usepgfplotslibrary{patchplots}
\usepackage{grffile} 
\usepackage{tikzscale}
 
\setlist[enumerate]{topsep=1ex,itemsep=-1ex,partopsep=1ex,parsep=1ex}
\setlist[itemize]{topsep=1ex,itemsep=-1ex,partopsep=1ex,parsep=1ex} 

\usepackage[unicode=true,bookmarks=true,bookmarksopen=false,
	bookmarksopenlevel=1,pdfborder={0 0 0},backref=false,
	colorlinks=true,allcolors=magenta,pdfstartview=Fit,
	pdfcenterwindow=true,pdfdisplaydoctitle=true,
	linktocpage=true]{hyperref}

\empheqset{box=\fbox}

\usetikzlibrary{calc,patterns,shapes}
\pgfplotsset{compat=1.15}
\tikzstyle{axis}=[->]
\tikzstyle{vector}=[-stealth, very thick]

\setlist[enumerate]{topsep=1ex,itemsep=-1ex,partopsep=1ex,parsep=1ex}
\setlist[itemize]{topsep=1ex,itemsep=-1ex,partopsep=1ex,parsep=1ex}
  
\hypersetup{
pdftitle={\doctitle\ -- \docsubtitle},
pdfsubject={\docversion},
pdfauthor={\docauthor},
pdfproducer={\docauthor},
pdfcreator={\docauthor}
}
 
\newcommand{\dd}{\mathrm{d}}
\newcommand{\pd}{\partial} 
 
\newcommand{\optsub}[1]{\ifthenelse{\equal{#1}{}}{}{_{\obj{#1}}}} % optional sub
\newcommand{\optsup}[1]{\ifthenelse{\equal{#1}{}}{}{^{\obj{#1}}}} % optional sup

\newcommand{\obj}[1]{\mathrm{#1}}

\newcommand{\tdd}[2][]{\ifthenelse{\equal{#1}{}}{\frac{\dd #2}{\dd t}}{\left.\frac{\dd #2}{\dd t}\right|_{\obj{#1}}}}
\newcommand{\tdddd}[2][]{\ifthenelse{\equal{#1}{}}{\frac{\dd^2 #2}{\dd t^2}}{\left.\frac{\dd^2 #2}{\dd t^2}\right|_{\obj{#1}}}}
\newcommand{\tddi}[2][]{\ifthenelse{\equal{#1}{}}{\dd #2/\dd t}{\left.\dd #2/\dd t\right|_{\obj{#1}}}} % inline
\newcommand{\tddddi}[2][]{\ifthenelse{\equal{#1}{}}{\dd^2 #2/\dd t^2}{\left.\dd^2 #2/\dd^2 t\right|_{\obj{#1}}}} % inline

	{\begin{inparaenum}[(\itshape i\/\upshape )]} {\end{inparaenum}}
 
\usepackage{multirow}
\usepackage{booktabs}
\usepackage{appendix}
\usepackage{ulem}
\usepackage{float}

\begin{document}
%----------------------------------------------------------------------

%---------------------------------------------------------------------- 
\title{
Analysis of collisional and facility effects in a magnetic nozzle plasma expansion} 
 
\author{Manuel Cortés-Hernán}
\email{manuelco@pa.uc3m.es}
\thanks{Corresponding author.}
\author{Mario Merino} 
\author{Diego García-Lahuerta} 
\author{Eduardo Ahedo} 
\affiliation{Department of Aerospace Engineering, Universidad Carlos III de Madrid, Legan\'es, Spain} 
%----------------------------------------------------------------------

%----------------------------------------------------------------------

\begin{abstract} 
An axisymmetric, quasineutral three-fluid model is proposed to study plasma expansion in a magnetic nozzle in the presence of neutrals coming either from the plasma source or as a homogeneous background. {Electron cooling in the plume is achieved by treating the electron heat flux as convective rather than conductive, eliminating the need to postulate anomalous cross-field or parallel diffusion of momentum or energy.} Solutions are presented for the high electron magnetization limit, in which the main electron flow variables can be integrated along magnetic field lines. Ionization, elastic, and charge-exchange collisions with neutrals do not alter the main qualitative features of the plasma expansion established in previous collisionless models. Ionization enhances the plasma flow in the nozzle and leads to additional electron cooling, which reduces the potential drop along the nozzle. Nozzle efficiency is quantified in terms of magnetic thrust gain and plume divergence angle. Two types of boundary conditions for the electron flow are evaluated: local throat current-ambipolarity conditions and global current-free conditions at the outer boundary (i.e., metallic vacuum chamber walls). The latter are shown to be physically more realistic: they incorporate the effect of the chamber walls on plasma expansion by shaping the ambipolar electric field and align better with experimental trends regarding background pressure.

\end{abstract}

\maketitle

%%%%%%%%%%%%%%%%%%%%%%%%%%
\section{Introduction}
%%%%%%%%%%%%%%%%%%%%%%%%%%

Electrodeless plasma thrusters (EPTs) have gained interest from the space propulsion community in recent  decades thanks to their theorized advantages over other well-established electric propulsion technologies \cite{ahed11s,bath17a}.
The lack of electrodes and neutralizing cathodes in EPTs offers possible benefits over conventional electric propulsion systems, such as extended operational lifetimes, fuel versatility \cite{shepp21a}, and broad thrust-to-specific impulse throttleability.
EPT devices such as the Helicon Plasma Thruster (HPT) \cite{char03b, bati09n, taka11b, nava18a} and the Electron-Cyclotron Resonance Thruster (ECRT) \cite{serc87, cann15b, inch23a} are now under development.
In these devices, plasma acceleration occurs in a so-called magnetic nozzle (MN) \cite{ande69, ahed10f, meri16g,kaga20a} that, similarly to a de Laval nozzle in chemical propulsion, converts (electron) thermal energy into (ion) directed kinetic energy.
MNs operate contactlessly thanks to the Lorentz force and, among other merits, offer the possibility of exerting thrust vectoring on the MN without moving parts \cite{meri17b}.

Seminal work by Ahedo and Merino \cite{ahed10f} analyzed the plasma expansion in a propulsive MN under in-space and collisionless conditions in the electron high-magnetization limit. Using an axisymmetric two-fluid model (DIMAGNO), they demonstrated the 2D character of the plasma expansion, explained the mechanisms for magnetic thrust generation via the electron azimuthal current, and discussed the relevance of boundary conditions. Their later works discussed plasma detachment via ion demagnetization \cite{meri14a}, ion and electron thermodynamics via polytropic relations \cite{meri15a}, and the influence of the plasma-induced magnetic field \cite{meri16b}.

However, the experimental conditions of a MN for an EPT operating in a laboratory can differ largely from the ideal conditions of that model. In particular, the existence of a background pressure in the vacuum chamber, $p_{bg}$, and the in-plume ionization of the gas emitted from the plasma source can affect the expansion of the plasma beam.
Vialis et al. \cite{vial17}, operating a small ECRT with Xe, found that thrust (and thrust efficiency) dropped significantly as $p_{bg}$ increased from 0.6 mPa to 1.3 mPa, although no clear reason for this was identified in their work. The divergence of the plasma plume also increased with $p_{bg}$, and the thruster floating potential decreased by 25\%. 
Caruso and {Walker} \cite{caru18}, testing an HPT with argon at low (1.6 mPa) and high (40 mPa) background pressures, found that, on the plume axis and under the highest pressure conditions, ion energies were tens of eV lower, electron temperatures a few eV lower, and higher plasma densities were observed.
Wachs and Jorns \cite{wach20a} analyzed background pressure effects on ion dynamics in the MN of a small ECRT operating with Xe. As  $p_{bg}$ increased from 0.1 to 3.5 mPa, the downstream ion velocity decreased by 20\%.
They suggested inelastic electron–neutral collisions in the plume as the mechanism responsible for the decrease in ion energy, which consumed up to 39\% of the incident power in the plume.
Desangles et al. \cite{desa23a} conducted additional tests with the ECRT of \cite{vial17}, with $p_{bg}$ ranging from 0.025 to 0.5 mPa, and the results showed a significant decrease in ion energy and thrust.

These experimental studies must be complemented with theoretical work that can, first, understand and quantify facility effects under diverse MN conditions and, second, extrapolate lab measurements to in-space conditions. Several models have addressed this task.   Andriulli et al. \cite{andr24} carried out a fully kinetic study of facility pressure effects on RF-source magnetic nozzles, based on a particle-in-cell (PIC) 2D model developed previously in \cite{andr22}. Background pressures up to 10 mPa were considered, and significant in-plume ionization was reported, leading to thrust losses of more than 13\% at high $p_{bg}$. Their model incorporates anomalous Bohm-type collisionality
{which effectively limits the maximum Hall parameter to about 64.} To reduce the huge PIC computational cost arising from the disparate time scales of the different plasma species, they used a vacuum permittivity augmented  713 times and an ion mass reduced 250 times, and claimed that these distortions did not affect the plasma response much. 

An alternative to MN full-PIC models is full-fluid,  quasineutral models, which are much cheaper computationally and thus more suitable for broad parametric investigations. However, fluid models require the pressure tensor to be isotropic and  a good closure of the fluid hierarchy, either at the momentum equation level (an example is a polytropic law, as in \cite{meri15a}) or at the energy equation level (i.e., an expression for the heat flux vector). Electrons are the most problematic species due to their large magnetization and being mostly a confined population. Marks et al. \cite{mark20a} used the 2D 3-fluid code Hall2De (originally designed for Hall thrusters) to analyze the MN measurements of \cite{corr19d} on the MN of an ECRT. The code solved a full electron energy equation with a Fourier-type heat flux vector.  
The authors found that this model over-predicted thermal conductivity, leading to a near-isothermal plume, in contrast with the significant plume cooling widely  reported in experiments \cite{corr19d, litt16,zhan16a}.  To reproduce that cooling, they added an isotropic anomalous electron resistivity (affecting the cross-field and parallel fluxes of momentum and energy) and fixed the downstream electron temperature.

{This work presents a 3-fluid MN model, which naturally extends DIMAGNO to a collisional plasma with highly magnetized electrons (i.e., very large values of the Hall parameter). 
The model incorporates an alternative closure for the electron energy equation, which extends the simple polytropic closures \cite{meri15a, arak16, cich17b}, and departs from conventional diffusive  heat flux closures \cite{mark20a, zhou22a}}.
The proposed closure is based on kinetic (Vlasov-based) studies suggesting  that the parallel electron heat flux in a MN rarefied plasma expansion is mainly convective \cite{mart15a, ahed20a, zhou21a}. {It will be shown that this choice leads to plume cooling naturally and recovers the polytropic law in the collisionless limit.} Additionally, the model applies global downstream conditions  (similar to \cite{andr24}),  instead of the local conditions of \cite{mark20a}. 
{Beyond the model formulation}, the main goals of this work are to analyze the effects of both in-plume ionization of neutral gas emitted  from the source and background pressure on MN performance. The influence of several MN parameters will also be investigated. Additionally, the electrical bridge between the thruster and the downstream region (i.e., the vacuum chamber walls or space vacuum), dependent on the choice of  boundary conditions, is examined.

Numerically, the new model (DIMAGNO-DG) is treated with a Discontinuous Galerkin zeroth-order finite element approach (equivalent to the finite volume method) developed recently by \cite{meri23a} to study the collisionless plasma expansion in a magnetic arch. 
This approach overcomes key limitations of the method of characteristics implemented in the original DIMAGNO model, notably the inability to simulate the peripheral space surrounding the main plasma jet and regions of subsonic ion flow, or to deal with shock-like structures. 
The model continues to take advantage of the very-high electron magnetization, a situation that may not hold if anomalous resistivity is added.

The rest of the paper is  organized as follows. Section \ref{sec:model} presents the three-fluid model. Section \ref{sec:vacuum_expansion} discusses the plasma expansion in vacuum, including specific subsections on performance and different boundary conditions. Section \ref{sec:collisionless_limit} details the collisionless limit for various inlet conditions. Section \ref{sec:back_pressure} analyzes the effects of background pressure. Section \ref{sec:model_validation} discusses the validity of the model, the effects of using a finite plume size, and the intrinsic 2D character of the MN expansion.
Finally, {Section} \ref{sec:conclusions} gathers the main conclusions of the study. 

%%%%%%%%%%%%%%%%%%%%%%%%%%
\section{Three-fluid plasma model} \label{sec:model}
%%%%%%%%%%%%%%%%%%%%%%%%%%

A plasma source of radius $R_0$ emits a plasma flow composed of electrons ($e$), singly-charged xenon ions ($i$) and neutrals ($n$), 
into an axisymmetric computational domain $\Omega$ of length and radius $L$. 
The plasma is channeled by a diverging, axisymmetric, magnetic field $\bm B (z,r)$, which admits a magnetic streamfunction $\psi(z,r)$, satisfying
\begin{equation}
    \nabla\psi=rB\bm 1_\perp.
%\bm B = r^{-1}\nabla \psi \times \bm 1_\theta.
\label{eq:B_psi_relation}
\end{equation}
The cylindrical and magnetic reference frames are $\{\bm 1_z,\bm 1_r,\bm 1_\theta\}$ and  $\{\bm 1_\parallel,\bm 1_\perp,\bm 1_\theta\}$, with $\bm 1_\parallel=\bm B/B$ and $\bm 1_\perp=\bm 1_\theta\times \bm 1_\parallel$ in the meridian plane. 
Although the study is applicable to any divergent configuration of a MN, the results are presented for the magnetic field of a single current loop flowing along $\bm 1_\theta$, and located at $(z,r) = (0,2R_0)$. Figure \ref{fig:sketch}(a) sketches the studied domain (dashed limits) and the relative magnitude of the magnetic field, with $B_0=B(0,0)$ proportional to the current loop and large enough to neglect effects of the self-induced magnetic field {within the simulation domain}.
%Additionally, the results are gathered for the zero plasma-beta parameter, $\beta = \mu_0nT_e/B^2 \ll 1$, thus neglecting the induced magnetic field with respect to the applied one.

%----------------------------------------------------------------------
\subsection{Ion and neutral equations}
%----------------------------------------------------------------------

The plasma is considered quasineutral and analyzed with a three-fluid model.
The continuity, momentum and energy fluid equations for  ions and neutrals  read
\begin{align}
&\partial_t n_i  + \nabla \cdot (n_i \bm u_i) = S_{ion},
\label{eq:ioncontinuity}
\\
&
\partial_t (m_in_i\bm u_i)  + \nabla \cdot (m_in_i\bm u_i\bm u_i +p_i \bm I )
= 
-en_i\nabla \phi + en_i\bm u_i \times \bm B  + \bm R_{in},
\label{eq:ionmomentum}
\\
&
\partial_t(n_i\mathcal{E}_i) 
+ \nabla\cdot[(\mathcal{E}_i + T_i)n_i\bm u_i] 
= 
-en_i\bm u_i\cdot\nabla\phi 
+Q_{in},
 \label{eq:ionenergy}
 \\
&
\partial_t n_n + \nabla \cdot (n_n \bm u_n) = -S_{ion},
\label{eq:neutralcontinuity}
\\
&
\partial_t (m_in_n\bm u_n)  + \nabla \cdot (m_in_n\bm u_n\bm u_n +p_n \bm I ) = - \bm R_{in},
\label{eq:neutralmomentum}
\\
&
\partial_t (n_n\mathcal{E}_n) 
+ \nabla\cdot[(\mathcal{E}_n+ T_n)n_n\bm u_n] 
= 
-Q_{in},
\label{eq:neutralenergy}
\end{align}
where, apart from standard symbols: $n_i=n_e$ is the quasineutral plasma density;
$\phi$ is the electrostatic potential; $p_\alpha=n_\alpha T_\alpha, \ (\alpha=i,n)$ are  pressures, 
$ \mathcal{E}_{\alpha} = m_iu_{\alpha}^2/2 + {3}T_{\alpha}/2 $ 
are total energies per particle, and $\mathcal{E}_{\alpha}+T_\alpha$ are the respective enthalpies. Ionization and charge-exchange collisions define the source terms
\begin{align} 
    &
    S_{ion}=n_e\nu_{ion},
    \qquad 
    S_{cex}=n_e\nu_{cex}, 
    \label{eq:Sion_Scex_def}
    \\
    &
    \bm R_{in} = m_i [S_{ion}\bm u_n - S_{cex} (\bm u_i - \bm u_n)], 
    \label{eq:Ri_def}
    \\ 
    &
     Q_{in}=S_{ion}\mathcal{E}_n-S_{cex}(\mathcal{E}_i-\mathcal{E}_n),
     \label{eq:Qi_def}
\end{align}
with frequencies $\nu_{ion}=n_nR_{ion}$ and $\nu_{cex}=n_nR_{cex}$. The ionization and charge-exchange rates, $R_{ion}(T_e)$ and $R_{cex}(|\bm u_i-\bm u_n|)$ are defined, together with other collisional parameters, in Appendix A of Ref. \cite{bell21a}.
For both ions and neutrals, the mechanical energy dominates over the thermal energy, so neglecting heat conduction is acceptable. 
Both $Q_{in}$ and $R_{in}$ omit a small contribution from electrons, proportional to $m_e$. 
The magnetization of the  heavy ions is often marginal.  
Since the plasma-induced magnetic field is neglected, the above set of equations is complete except for the electric potential $\phi$. The electron equations close the plasma model.

%----------------------------------------------------------------------
\subsection{Model for highly magnetized electrons} \label{subs:electron}
%----------------------------------------------------------------------

Neglecting electron inertia and assuming quasineutrality, the set of electron equations is
\begin{align}
&\partial_t n_e +\nabla \cdot (n_e \bm u_e) = S_{ion},
\label{eq:electroncontinuity}
\\
&\bm 0 = -\nabla p_e + en_e \nabla \phi - en_e \bm u_e \times \bm B - \nu_e m_e n_e  \bm u_e,
\label{eq:electronmomentum}
\\
&
\partial_t(n_e\mathcal{E}_e)+
\nabla\cdot[(\mathcal{E}_e\bm+T_e) n_e\bm u_{e}] = en_e \bm u_e \cdot \nabla \phi - S_{ion}\mathcal{E}_{inel},
\label{eq:electronenergy}
\end{align}
where: $\nu_e$ is  the effective electron momentum collision frequency, 
accounting {only} for elastic collisions with neutrals and ions, and the velocities of these heavy species have been omitted compared to $\bm u_e$; 
$ \mathcal{E}_e = {T_e}/{(\gamma_e-1)}$
is the internal energy per electron, 
with $\gamma_e$ the specific heat ratio of the electron fluid; and $\mathcal{E}_{inel}$ is the inelastic loss per electron due to ionization and excitation/de-excitation processes (defined together with $\nu_e$ in  Ref. \cite{bell21a}).  {The mobility and Hall parameter are then defined as
\begin{equation}
    \mu_e=e/m_e\nu_e,
    \qquad 
    \chi=B\mu_e.
    \label{eq04}
\end{equation}
}

The adopted model for the electron energy equation \eqref{eq:electronenergy} differs from the one used in the Hall2De \cite{mark20a} and HYPHEN \cite{zhou22a} codes. These two  assume the electron fluid to be a simple gas with $\gamma_e=5/3$ and a conductive heat flux vector, $\bm q_e=-\bar{\bar K}\cdot \nabla T_e$ , with $\bar{\bar K}$ a thermal conductivity tensor. However,  the application of this model shows that the conductive parallel heat flux is dominant in the plume, making it near-isothermal, unless anomalous energy diffusion is included \cite{mark20a,pera25a,fern26b}. On the other hand, kinetic investigations on MNs \cite{ahed20a,zhou21a}, and other rarefied plasmas \cite{stan10,malo75,bell85}, demonstrated, first,
that the electron population is a mixture of two subpopulations (confined and free electrons) with different dynamics and, second, that the electron energy transport is mainly convective. Those investigations conclude that the mixture would be reasonably well characterized with a specific enthalpy $(\mathcal{E}_e + {T_e})=T_e\gamma_e/(\gamma_e-1)$.
{Furthermore, it will be shown that Eq. \eqref{eq:electronenergy}  extends naturally to the weak-collisional regime} the simple polytropic relation 
\begin{equation}
    T_e\propto n_e^{\bar \gamma_e-1}
    \label{eq01}
\end{equation}
for $T_e$ \cite{meri15a, arak16}, an empirical law aimed to recover the  electron cooling observed experimentally, with $\bar \gamma_e$ in the range 1.1-1.3 generally, {and much used in plume codes} \cite{tacc14,cich17b}.

The present study focuses on  collisional effects on the stationary plasma response, while remaining in the limit of very high electron magnetization, i.e., $\bar \chi\ll 1$,
where $\bar\chi=1/\chi$ is the inverse of the Hall parameter.
Figure \ref{fig:sketch} (b) shows that the Hall parameter for the simulation with the highest collisionality in this paper satisfies that criterion. 
Then, neglecting the temporal terms in Eqs. \eqref{eq:electroncontinuity} and \eqref{eq:electronenergy},
the equations for electron continuity and internal energy, up to dominant order, reduce to
\begin{align}
    &
    B\ \nabla_\parallel  \frac{n_e u_{\parallel e}}{B}
     \simeq S_{ion},
     \label{eq:electroncontinuity_red}
     \\
     &
    B\ \nabla_\parallel \Big(\frac{\gamma_e T_e}{\gamma_e-1}\frac{n_e u_{\parallel e}}{B}\Big) 
    \simeq 
    u_{\parallel e} \nabla_\parallel p_e 
    - S_{ion}\mathcal{E}_{inel}
    +\nu_e m_e n_e u_e^2,
    \label{eq:electronenergy_red}
 \end{align}
while  the components of the momentum vector equation \eqref{eq:electronmomentum} become
\begin{align}
&
    0=-\nabla_\parallel p_e+en_e\nabla_\parallel\phi-m_e\nu_e n_e u_{\parallel  e},
     \label{eq:lectronmomentum_red}
    \\
    &
    u_{\theta e}=\frac{1}{1+\bar \chi^2 }
    \Big(
    \frac{\nabla_\perp \phi}{B}
    -\frac{\nabla_\perp p_e}{en_eB}
    \Big),
     \label{eq:ue_theta}
    \\
    &
    u_{\perp e}=\bar \chi u_{\theta e}.
     \label{eq:ue_perp}
\end{align}
Since $n_e = n_i$, this set of 5 equations determines $\bm u_e$, $T_e$, and $\phi$.

Now, instead of $u_{\parallel e}$, $T_e$, and $\phi$, it  is convenient to use the variables
\begin{equation}
    G=\frac{n_eu_{\parallel e}}{B},
    \qquad
    A=\Big(\frac{T_e}{T_{e0}}\Big)\Big(\frac{n_{e0}}{n_e}\Big)^{\gamma_e -1},
    \qquad
    \Phi=\phi-\frac{\gamma_e}{\gamma_e-1}\frac{T_e}{e},
    \label{eq:electron_properties}
\end{equation}
which respectively represent the electron flow, the adiabaticity function, and the thermalized potential; the subindex notation  $n_{e0}=n_e(0,0)$, $T_{e0}=T_e(0,0)$, and so on for other variables at the origin, is used.
Operating with Eqs. \eqref{eq:electroncontinuity_red}-\eqref{eq:lectronmomentum_red} we end up with 
\begin{align}
&
    \nabla_\parallel G=\nu_{ion}n_e/B,
    \label{eq:G_integration}
    \\
&
   \nabla_\parallel \ln A
   =
   \frac{\gamma_e-1}
   {T_e u_{\parallel e}}\Big[ 
   \nu_e m_eu_e^2-\nu_{ion}\Big(\frac{\gamma_e}{\gamma_e-1} T_e+\mathcal{E}_{inel}\Big)
   \Big],
   \label{eq:C_integration}
   \\
   &
    \nabla_\parallel \Phi
   =\frac{1}{eu_{\parallel e}}
   \Big[
   \nu_{ion}\left(\frac{\gamma_e}{\gamma_e-1} T_e+\mathcal{E}_{inel}\right)
   -\nu_e m_e u_{\theta e}^2
   \Big],
   \label{eq:Phi_integration}
\end{align}
which determine the variations of the three variables along each magnetic line. Once the $(z,r)$ maps of these 3 functions are computed, the maps of $u_{\parallel e}$, $\phi$, and $T_e$ are readily obtained.

%%%%%%%%%%%%%%%%%%%%%%%%
\subsection{Full plasma model}
%%%%%%%%%%%%%%%%%%%%%%%%

The plasma model is now complete. Prior to integrating the set of ion and neutral temporal-spatial equations \eqref{eq:ioncontinuity}--\eqref{eq:neutralenergy}, it is convenient to  write them in fully conservative form. This requires eliminating the terms involving $\nabla\phi$ on the right-hand sides of Eqs. \eqref{eq:ionmomentum} and \eqref{eq:ionenergy}. For the ion momentum equation, the electron momentum equation \eqref{eq:electronmomentum} is added to it. For the ion energy equation,  the following substitution is performed, 

\begin{equation}\label{eq:Ei_rearrange}
    n_i\bm u_i\cdot\nabla\phi=\nabla\cdot (\phi n_i\bm u_i)- \phi (S_{ion}-\partial_t n_i).
\end{equation}

After applying the rearrangement, the system is expressed almost entirely in a conservative form, except for a residual term, namely $\phi\partial_tn_i$. This contribution can be neglected, as the steady-state solution is ultimately sought.

Appendix \ref{sec:nmerical_integration} explains the details of the Discontinuous Galerkin finite element  method implemented in DIMAGNO-DG to integrate the present model and which was first presented in \cite{meri23a} for the collisionless limit, when $G$, $\Phi$, and $A$ are constant along each magnetic line. 
In contrast, the original DIMAGNO did not include collisions and was based  on the method of characteristic lines to integrate the ion equations \cite{ahed10f}. While efficient, this limited
the integration to only the main plasma plume, bounded by the magnetic line departing from $(z,r)=(0, R_0)$.That model was also limited to polytropic-like laws \eqref{eq01} for the electron temperature with $\bar \gamma_e$ equal or near-equal to 1, to avoid the merging of characteristic lines when ions became hypersonic. 
The equations of the heavy species are integrated on a triangular mesh of 80,000 elements formed by dividing squares with a facet size length of 0.9 mm along their diagonal. An interpolation routine is then required between this mesh and the magnetically aligned results for the electron variables. 
The solution procedure includes an iterative scheme to compute the functions $G$, $A$, and $\Phi$ in equations \eqref{eq:G_integration}--\eqref{eq:Phi_integration}, which starts off from the collisionless solution.

%%%%%%%%%%%%%%%%%%%%%%%%
\subsection{Boundary conditions in the nominal simulation}
%%%%%%%%%%%%%%%%%%%%%%%%

In the axisymmetric domain $\Omega$ of the MN, Fig. \ref{fig:sketch} (a), three types of boundaries are distinguished: the symmetry axis, which is crossed by neither magnetic lines nor species flows; the MN throat (subindex  0), at $z=0$, with plasma entering  the domain; and the outer boundary  (subindex  $P$) with plasma \textit{leaving} the domain (both downstream, laterally, and backward).

The 'nominal' simulation N to be discussed first 
describes the expansion of an EPT exhaust with incomplete utilization, i.e., featuring a flow of neutrals as well as a flow of plasma, expanding into {vacuum.}  
The simulation has the following boundary conditions (BCs), at the throat $z=0$:
\begin{align}
&
\phi(0,r) = 0,
&
\label{eq:phi_inlet}
\\
&
n_\alpha(0,r) = n_{\alpha 0}(10^{-3(r/R_0)^2} + \epsilon)/(1+\epsilon), 
&
\quad \alpha=i,n,
\label{eq:n_inlet}
\\
&
T_{\alpha} (0,r)  = T_{\alpha 0}, 
&
\quad \alpha=e,i,n,
\label{eq:T_inlet}
\\
&
u_{z\alpha}(0,r) = M_{\alpha 0} c_{\alpha 0},  
&
\quad \alpha=i,n,
\label{eq:ualpha_inlet}
\end{align}
with 
\begin{equation*}
\label{eq:sonic_vel}
    c_{i}=\sqrt{\frac{\gamma_e T_e+(5/3)T_i}{m_i}},
    \qquad
    c_{n}=\sqrt{\frac{5T_n}{3m_i}},
\end{equation*}
and $M_{\alpha 0}$ axial Mach numbers.
For these boundary conditions, the ion and neutral mass flows are
\begin{equation}\label{eq:dot_m_alpha}
    \dot m_{\alpha 0}\simeq 0.145\  \pi R_0^2 \ m_i u_{\alpha 0}n_{\alpha 0},
\end{equation}
so that  $\bar{n}_{\alpha0}=0.145 n_{\alpha_0}$ is the radially averaged density at the MN throat. It is also found that 95\% of the ion mass flow is contained within the tube $r\leq 0.722 R_0$.

{A numerical difficulty of the present  stationary, very-low-$\bar \chi$  electron model is the simulation of  the highly rarefied region around the coil and external to the magnetic lines emanating from the MN throat. This region, still quasineutral,  is filled with plasma during the MN transient and through electron cross-field diffusion. Both phenomena are absent here, and indeed DIMAGNO excluded that region. To include it within our stationary model the injection surface has been extended up to $r=2R_0$ and a lower bound on the plasma density $\epsilon n_{e0}$, with $\epsilon\ll 1$, has been imposed at the surface of that highly-rarefied  region. The nominal simulation takes $\epsilon= 10^{-3}$, based on results from \cite{andr24},  and the sensitivity of the solution to $\epsilon$ is analyzed later.}

To complete the above set of  BCs, an additional condition is needed on the parallel electron velocity (or current). This can be set at either the throat or the outer boundary. In the first and simplest case, the Throat local Current Ambipolarity (TCA) condition 
\begin{equation}
   u_{\parallel e}(0,r)\equiv u_{ze}(0,r) =u_{zi}(0,r)
\label{eq:ue_inlet} 
\end{equation}
is imposed. Other possibilities for setting $u_{\parallel e}$ are discussed in the Section \ref{subsec:OFW}.

%%%%%%%%%%%%%%%%%%%%%%%%
\section{Plasma expansion into vacuum}\label{sec:vacuum_expansion}
%%%%%%%%%%%%%%%%%%%%%%%

\subsection{2D response for the nominal simulation}

Numerical values for the variables of the above BCs and other simulation parameters of case N are listed in Table \ref{Table:Initial_conditions}. These were mainly selected from the prototypes and the experimental data reported in \cite{wach20a,inch23a}.
Notice that at the MN throat: a propellant utilization of 62.5\% leaves margin for ionization within the MN; 
72\% of the thrust and 97\% of the power are still provided by the electron population; 
and the contribution of neutrals to thrust and power is negligible. 
The selection of a subsonic ion Mach number, $M_{i0}=0.5$, is based on ensuring that  $\phi(z,0)$ decreases monotonically from the throat: indeed,  for simulation N, 
higher values of $M_{i0}$ would lead to a non-monotonic behavior of $\phi(z,0)$ with a local maximum after the throat, which locally reduces the ion velocity. This behavior, which we consider artificial, is also observed, for instance, in \cite{andr24} when a sonic ion flow is used at the throat.
Furthermore, several experiments indicate that the sonic transition occurs downstream of the MN  throat \cite{corr19d,vinci22c, coll19}.
It will be shown later that  a lower plasma collisionality at the throat supports a higher value of $M_{i0}$ while keeping  $\phi(z,0)$ monotonic.

Figures \ref{fig:electronic_properties_2d} and \ref{fig:ref_properties_2d} show 2D maps of relevant plasma variables for simulation N.
Figure \ref{fig:electronic_properties_2d} depicts the dimensionless ratios of the electron functions $G$, $\Phi$, and $A$ characterizing the evolution of electrons along the magnetic lines. The high electron collisionality around the throat is caused by $n_n$ and explains the steep axial gradients of those functions in that region. Moving  downstream, as $\nu_e$ and $\nu_{ion}$ in Eqs. \eqref{eq:G_integration}-\eqref{eq:Phi_integration} decay, the above electron functions tend to asymptotic values, which are \textit{different} on each magnetic line. For instance, at on-axis downstream point D, defined by $(z,r)_D=(L, 0)$, these values are  $G/G_0 \simeq 1.41$, $A/A_0 \simeq 0.616$, and $e(\Phi-\Phi_0)/T_{e0} \simeq 0.134$.

Turning now to Fig. \ref{fig:ref_properties_2d}, the following features are observed.  The map of $n_n$ shows that neutrals expand almost hemispherically, with additional depletion caused by ionization. In contrast, the map of $n_e(z,r)$ indicates that ions and electrons are well confined within the magnetic lines, and their geometric expansion is limited to central magnetic streamsurfaces.  
The behavior of $T_e(z,r)$ corresponds  to an adiabatic expansion, with the correction factor given by function $A$. Similarly, along each magnetic line, $\phi$ approximately follows the generalized Boltzmann relation of Eq. \eqref{eq:electron_properties}, adjusted  for the small variation of $\Phi$. The equipotential lines have lobed shapes due to a line of local minima emanating approximately from the edge of the thruster exit, which is analyzed below.

For velocities and current densities, it is convenient to separate the azimuthal component from the vector component in the meridian plane: for instance, $\bm j_e=\bm{\tilde \jmath}_e+ j_{\theta e}\bm 1_\theta$, and so on. 
In Fig. \ref{fig:ref_properties_2d} we first observe the electron currents in the meridian plane, $\bm {\tilde\jmath}_e(z,r)$, which, due to the high magnetization regime, follow the magnetic lines. In contrast, $\bm {\tilde\jmath}_i(z,r)$ shows that ions rapidly detach inwardly from the magnetic lines, due to their supersonic expansion away from the throat \cite{meri14a}.
The differing behaviors of ion and electron meridian currents imply that the current ambipolarity condition, Eq. \eqref{eq:ue_inlet}, is broken almost immediately, and  
$\bm {\tilde\jmath}(z,r) = \bm {\tilde\jmath}_i + \bm {\tilde\jmath}_e$ 
develops loops of electric current in the meridian plane (while  the plasma beam remains current-free across a surface totally enveloping the injection region). 
The azimuthal electron current (with $j_{\theta e}= -en_eu_{\theta e}$) is negative, meaning it is diamagnetic and provides positive magnetic thrust. There is also a small azimuthal ion current, $j_{\theta i}$, which is paramagnetic \cite{ahed11d}, and reduces the magnetic thrust by about 1 \%.
The map of the magnetic force $(-j_{\theta e}B_r)$ shows that the differential magnetic thrust is concentrated within a few radii downstream of the main plasma plume, though residual thrust still remains at the end of the numerical domain. Notably, both $u_{\theta e}$  --Eq. \eqref{eq:ue_theta}-- and  $j_{\theta e}$ scale as $1/B_0$, 
whereas the force density $(-j_{\theta e}B_r)$ is essentially independent of $B_0$ (400 G in this case).

The bottom row of Fig. \ref{fig:ref_properties_2d} shows the three collision frequencies included in the plasma model. For simulation N, up to a few radii downstream, these roughly satisfy the following ordering:
\begin{equation}\label{eq:collisional_freq}
    \nu_e=O(10^6 s^{-1})\gg \nu_{ion}=O(10^4 s^{-1})\gg \nu_{cex}=O(10^3 s^{-1}). 
\end{equation}
This implies that ionization and charge-exchange collisions contribute comparably to the ion-neutral collisional force $\bm R_{in}$ of Eq. \eqref{eq:Ri_def}.

%%%%%%%%%%%%%%%%
\subsection{Magnetic nozzle performance}
%%%%%%%%%%%%%
 
Following Appendix \ref{AppB}, the variations of ion mass flow $\dot m$, thrust $F$, and plasma power $P$ between the throat (0) and the  outer surface (P) are 
\begin{equation}\label{eq:thrust_mass_P}
    \dot m_{iP}-\dot m_{i0}=m_i\int_{\Omega} \dd \Omega\ S_{ion},
    \quad
\end{equation}
\begin{equation}
    F_P-F_0=\int_{\Omega} \dd \Omega \ (-j_\theta B_r),
    \quad
\end{equation}
\begin{equation}\label{eq:P_P}
    P_P-P_0=-\int_{\Omega} \dd\Omega\ S_{ion}\mathcal{E}_{inel},
\end{equation}
each expressed as a volume integral over $\Omega$.
The  thrust efficiency, including the plasma source contribution, then satisfies 
\begin{equation}\label{eq:efficiency}
    \eta= \dfrac{F_P^2}{2 \dot{m} P_{tot}}  = \frac{P_0}{ P_{tot}} \frac{F_0^2}{2 \dot{m} P_{0}}\frac{F_P^2}{F_0^2},
    %= \kappa_P \kappa_{MN}\kappa_F^2
\end{equation}
where $P_{tot}=P_0+P_{wall}+P_{inel}$ is the total deposited power into the plasma source, and $P_{wall}$, $P_{inel}$ denote, respectively, the wall losses and the electron inelastic losses (ionization and excitation) inside the source.
The present MN model does not address $P_0/P_{tot}$; it treats the quantities in ${F_0^2}/{2 \dot{m} P_{0}}$ as boundary conditions and determines only the last factor, $({F_P}/{F_0})^2$, the square of the thrust gain across the MN.

Table \ref{Table:End_values_reordered3} details the plasma performance in the MN for the various simulations studied. Its second column corresponds to simulation N under TCA conditions. In-plume ionization in the MN raises the ion mass flow by 18 percentage points, bringing the final propellant utilization to 74\%, while ionization and excitation losses consume 11\% of the plasma power. The magnetic force contributes to performance, producing a thrust gain of 45\%, equivalent to an efficiency increase by a  factor of 2.1. Although part of this thrust gain is due to the increased  propellant utilization by newly created ions, most of it reflects the increase of ion velocity in the plume: $u_{zi}(L,0)/u_{zi}(0,0)\simeq 4.6$. 
Together with the thrust gain originated by the magnetic force, the ambipolar electric field transfers 'electron' thrust and energy to 'ion' thrust and energy \cite{ahed11d}. By the outer boundary, only 4.8\% of the beam thrust and 10.3\% of its power remain in the electron fluid.
%which means that the plasma expansion is not complete yet. 
A further performance metric is the semiangle of the plume divergence, measured as $\alpha_{div}=\text{atan}(R^*/L)$, where  $R^*$ is the downstream radius of the streamtube enclosing 95\% of the plasma mass flow. 
For the nominal simulation, it takes a moderate value of $\alpha_{div}\simeq $34.3 deg.

%%%%%%%%%%%%%%%%%%
\subsection{Outer boundary closure for the electron model} \label{subsec:OFW}
%%%%%%%%%%%%%%%%%%

Although the TCA condition is the simplest to integrate numerically, it lacks solid physical justification, for two reasons. First, simulations of the whole system (source + MN) do not suggest local current ambipolarity at the exit \cite{meri16a, zhou22a}; the plasma is more collisional inside the source, and thus more prone to sustain $\bm{\tilde \jmath}\neq 0$ than within the MN itself. In fact, we already showed that current ambipolarity is violated immediately after the throat.
Second, the TCA condition provides no mechanism to electrically bridge the plume boundary surface P to a vacuum chamber wall or the undisturbed space vacuum \cite{domi25a}. This link must be established by determining electrical conditions at the outer boundary surface.

To illustrate the situation and define the outer BCs, let us consider the boundary P of the quasineutral simulation domain as adjacent to a conducting wall W, biased to a floating potential $\phi_W$, yet to be determined. 
Let  $\bm 1_n$ be the outward-pointing local unit vector normal to P. At any point Q of the quasineutral boundary surface P, a non-neutral Debye sheath develops, connecting point Q to the wall. The local potential drop across the sheath, $\phi_{W Q}=\phi_Q-\phi_W$ (assumed positive), governs the electron current reaching the wall, which satisfies \cite{pera22b}
\begin{equation}
    j_{neQ}\equiv(\bm j_{e}\cdot \bm 1_n)_Q =-en_{eQ}\sqrt{T_{eQ}/(2\pi m_e)}\exp(-e\phi_{W Q}/T_{eQ}).
    \label{eq02}
\end{equation}
The global current-free condition at boundary P requires  $\phi_W$ to satisfy the integral condition \cite{domi25a}
\begin{equation}
    \int_{P} dS [j_{niQ}+j_{neQ}(\phi_W)]=0
\end{equation}
where $j_{niQ}\equiv\bm (\bm j_{i}\cdot \bm 1_n)_Q$ does not depend on $\phi_W$. 
Once this wall potential is determined, $j_{neQ}$ is known at each point of the  boundary P, and $u_{\parallel eQ}$ satisfies
\begin{equation}
    u_{\parallel eQ}=
    \frac{\sqrt{T_{eQ}/(2\pi m_e)}
    \exp(-e\phi_{W Q}/T_{eQ})}{(\bm 1_\parallel\cdot \bm 1_n)_Q}.
\end{equation}
This Outer-boundary Floating  Wall (OFW) condition is proposed to substitute the TCA condition \eqref{eq:ue_inlet}, allowing Eq. \eqref{eq:G_integration} for $G$ to be integrated inwardly from the outer boundary toward the throat.

Figure \ref{fig:boundaries_1d} compares, for  simulation N,
the use of TCA and OFW conditions on the electric potential and currents along all the domain boundaries (except for the axis). The abscissa is the arc length variable $s$, advancing along the boundary P, with $s=0$ at $(z,r)=(0,0)$. 
All the plotted variables differ moderately between cases TCA and OFW, while  
$j_{ne}$ changes drastically at the MN throat.
In the OFW case,   $j_{ne}$ along P  defines $j_{\parallel e}$ at P, and the backward integration along each magnetic line departing from P to the throat yields the highly variable profile of  $j_{ne}$ at the  throat, where it  even changes sign and contrasts sharply with the constant behavior of the TCA case.
For both TCA and OFW cases, the electric potential at P is between -40 and -30 V. However, only using the OFW condition the wall potential ($\phi_W\simeq -53.6$ V) can be determined; it  is about 5-6 times $T_{e0}/e$, in agreement with kinetic modeling (for xenon) \cite{ahed20a}. 
The electron temperature along the outer boundary varies in the range 1.5-3.5 eV.

Table \ref{Table:End_values_reordered3} tells us that, in spite of the differences on $\bm {\tilde\jmath}_e$ 
for the TCA and OFW cases of the N simulation, the differences in the MN performance are minor. For the OFW case,
ionization is a bit stronger, explaining the larger power losses. However, the thrust gain is a bit lower, likely due to a slightly higher plume divergence  ($\alpha_{div}\simeq 35$ deg). 

The electron BCs at P applied  by {Marks} et al. \cite{mark20a} to their fluid model
set $j_n=0$ and fix the value of $T_e$  at P. The first one applies to a dielectric wall (which is not the common  case in a vacuum chamber and  does not permit an extension to uniform conditions in space vacuum) and the second one is not the most natural one.  Indeed, the imposition of $T_e$ at P plus the \textit{ad hoc} inclusion of \textit{parallel} anomalous resistivity  are due to the dominance of energy diffusion in the electron energy equation that tends to create an isothermal plume. The same problem was identified using the electron model of  HYPHEN and was recently discussed  in \cite{fern26b}.

%%%%%%%%%%%%%%%%%%%%%%%%%%%%%%%%%%%%%%
\section{The collisionless plasma expansion}\label{sec:collisionless_limit}
%%%%%%%%%%%%%%%%%%%%%%%%%%%%%%%%%%%%%

To understand the differences between the effects on performance of the plasma injected into the MN and the one created within it, a collisionless simulation B0, identical to simulation N but canceling all collisional effects involving the ejected neutrals (i.e.,  $\nu_{ion}$, $\nu_e$, and $\nu_{cex}$ are taken zero), was set up.  The numerical integration simplifies much since (a) the equations of neutrals are not needed, and (b) the electron-related functions $G$, $\Phi$, and $A$,  remain constant along each magnetic line, as already found in \cite{ahed10f,meri15a}.  
This model also  showed that $u_{\theta e}/r$ is constant along each magnetic line,  $u_{\parallel e}=G B/n_e$ is decoupled from the rest of the equations, and differences between TCA and OFW conditions affect  $u_{\parallel e}$ \textit{exclusively}. 
In addition, the collisionless simulation B0 allows us a direct comparison of the present model with the original DIMAGNO \cite{ahed10f}. 

The comparison between simulations N and B0 finds only mild differences  in the 2D maps of plasma variables.  To illustrate them, 
Figure \ref{fig:basic_ideal_properties} plots 3 maps for simulation B0-TCA. Compared to simulation N, $T_e(z,r)$ near the throat decreases less in case B0, since inelastic losses are absent; and the same lobes on equipotential lines are found.  
The plot of  $\bm{\tilde \jmath}_i$  adds two colored solid lines, corresponding to 
95\% flow surfaces, both departing from $(z,r)/R_0=(0, 0.722)$:
the red line corresponds to ions, and the black one to electrons and to a magnetic streamline. 
The separation between them illustrates clearly that ion detachment in the MN is progressive, it is already well developed at $z=2R_0$, and ions end with a near-conical expansion, without the need to break plasma quasineutrality \textit{anywhere} \cite{meri14a}.

Figure  \ref{fig:basic_ideal_properties} confirms that  the lobed equipotential lines of simulation N are not due to in-plume ionization.  There is ample experimental \cite{char10, litt19, coll19} and numerical \cite{chen20a} evidence of this behavior. The line of local minima of $\phi$ follows approximately the magnetic line 
departing from the thruster mouth edge, that is, the one separating the magnetized plasma beam from the isolated near-vacuum region.
In those works, the behavior of $\phi$ is  attributed  to warm ions with large lateral velocity components moving laterally outward and trying to cross the plume-vacuum interface. This would generate a positive electric charge buildup resulting  in a rise of electric potential to bind the fleeing ions to the magnetized electrons. 
We do not question that explanation, but it does not fit with the conditions of our expansion, where \textit{near-cold} ions 
are emitted \textit{axially} and the whole solution (even beyond the plasma-vacuum separatrix) is \textit{quasineutral} and thus the electric field \textit{ambipolar}. 
We propose therefore an alternative explanation.
Under the electron massless assumption, the ambipolar electric field responds immediately to maintain the electron momentum balance. For illustration, let us consider the radial equilibrium at a given $z=$const section: 
\begin{equation}\label{eq:grad_phi_r}
     \nabla_r \phi
     =
     u_{\theta e }B_z
     + 
     \frac{\nabla_r p_e}{e n_e}.
\end{equation}
Under usual conditions at the thruster exit,  the confining magnetic force and the expanding pressure force are dominant near the throat, so that the electric force is marginal (which implies that the electron  $E\times B$ azimuthal drift is marginal too). Therefore, the sign of the electric force is very sensitive to small variations in the other two forces.  {One of the parameters influencing it is the residual density $\sim \epsilon n_{e0}$ in the highly rarefied region around the coil, which is an input of our model.} Figure  \ref{fig:phi_study} illustrates this with radial profiles at $z=R_0$, for simulation B0  and three values of  $\epsilon n_{e0}$. The electric force is {one order of magnitude lower} than  the other two, and its value in the interval $0.5<r<2.5$ is much affected by the details of the decay profile of the pressure and magnetic forces. {The results indicate that the  development of a minimum of $\phi$ is favored by a lower residual density.  Beyond this local effect,  the value of $\epsilon$ has no effect on the MN performances.}

Table \ref{Table:End_values_reordered3} compares performances in simulations N and B0. For case B0, there is no increase in propellant utilization and thus no related power losses.  Without contribution from new ions in simulation B0, the thrust gain is 41\%, the downstream  ion velocity at the axis increases to $u_{zi}(L,0)/u_{zi}(0,0)\simeq 5.3$, and the plume divergence semiangle decreases slightly to 33.7 deg. 
The contribution of the in-plume ionization of neutrals to $F_P^2/F_0^2$ in simulation N is not negligible, enhancing $\eta$ by about 5-10\% (but at the expense of the electron power spent in those inelastic collisions).

Even so, the collisionless case is simpler to integrate and provides a good qualitative description and a fair quantitative one of the plasma response in the MN. Taking advantage of this numerical simplicity, two new collisionless simulations were run. The first one has $B_0=$ 100 G, and the performance was found  to be the same as for 400G, within a margin below 2\%. The only relevant change is that $j_{\theta e}$ quadruples, so $j_{\theta e}B_r$  and thrust $F_P$ remain identical.
This  result confirms that the main role of the magnetic field is to channel the electron flow. A lower magnetic field is beneficial for ion detachment, as it can occur earlier in the expansion \cite{meri14a}, but it lowers the electron magnetization too. As we will see in a later section, this reduces the extension of the plume where the present model is valid. 

The second additional simulation, B0m in Table   \ref{Table:End_values_reordered3}, considers a sonic ion flow at the entrance, $M_{i0}=1$, to assess the influence of this parameter on the  plasma response. Related changes in the throat BCs of Table \ref{Table:Initial_conditions} are as follows: ion velocity $u_{zi0}$ doubles and $\bar n_{i0}$ halves, so $P_0$ is 4.85 W instead of 4.53 W,  $F_0$ is 546 $\mu$N instead of 644 $\mu$N, and now $P_{i0}>P_{e0}$ and $F_{i0}>F_{e0}$. The plasma response (not plotted) shows a gentler decay in density and electrostatic potential near the throat, due to the higher ion momentum.  The lobes in the maps of $\phi$ persist. Table \ref{Table:End_values_reordered3} compares the performance in both cases \textit{for the selected throat BCs}:  although most of the downstream variables are rather similar in simulations B0 and B0m, the MN efficiency gain is 7\% larger, which highlights the importance of coupling the source and nozzle regions for an accurate quantitative assessment.

%%%%%%%%%%%%%%%%%%%%%%%%
\section{Effects of background pressure}\label{sec:back_pressure}
%%%%%%%%%%%%%%%%%%%%%%%

The presence of a background pressure, $p_{bg}$, in the beam expansion is modeled as a uniform population of neutrals at rest with temperature $T_n=290$ K (= 0.025 eV) and density $n_n=p_{bg}/T_n$. Although the physical model can include the combined effects of  background neutrals and those emitted by the source, we have opted to present simulations with the presence of background neutrals only, in order to highlight clearly the  effects of  $p_{bg}$. Hence, B0 is the simulation of reference in this Section,  and we discuss two new simulations, B2 and B4, corresponding to $p_{bg}=$  2 and 4 mPa, respectively; for comparison, the  experiments of \cite{wach20a} consider values of $p_{bg}$ up to 3.45 mPa.

With the adopted model assumptions, the background density affects only the ionization and electron-momentum collision frequencies $\nu_{ion}$ and $\nu_e$; charge-exchange collisions are considered marginal in these cases and thus neglected. For each simulation, both TCA and OFW boundary conditions are considered. 

Figure \ref{fig:electronic_properties_1d} shows the 1D profiles for the  three electron-related functions along the MN axis  for both the TCA and OFW cases. In contrast to what is observed for simulation N, where functions $G$, $\Phi$ and $A$ change steeply near the throat and then approach asymptotic values downstream (when the semispherical expansion of neutrals makes $n_n$ very low), simulations B2 to B4 present a gentle but persistent  variation of those functions.
This is of course related to the profile of $n_n (z,0)$, which is constant and equal to $0.5\cdot 10^{18}$m$^{-3}$ and $10^{18}$m$^{-3}$ for B2 and B4, respectively, while it decreases several orders of magnitude along the MN for simulation N, as shown in  Fig. \ref{fig:properties_1d}.
The trends of the 3 electron functions depend mainly on $\nu_{ion}\propto n_n$, except for the slight downstream increase  of function $A$ in simulation B4, caused by a higher  $\nu_e$.
The gradients of the 3 functions are a bit larger for the TCA cases than for the OFW ones,  due to the differences in $u_{\parallel e}$ between both cases.

The first row of Fig. \ref{fig:properties_1d} depicts the profiles along the MN axis of $n_e$, $T_e$ and $\phi$ for  different $p_{bg}$ (and simulation N) and the OFW case; trends are similar for the TCA case. As  $p_{bg}$ increases from B0 to B4, so does ionization. As a consequence,  $T_e$ decreases more rapidly, and the gradients of $\phi$  and $u_{zi}$ (in the second row) are gentler. The behavior of $n_e$  is the combined result of the beam  geometrical expansion and the lower gradient of  $u_{zi}$. The increase of the ion mass flow is proportional to $\nu_{ion}$,  and the gain in magnetic thrust is not much influenced by $p_{bg}$, since the decrease in ion velocity is mostly compensated  by the increase in ion mass flow. The presence of a residual magnetic force still at $z=L$ , observed in Fig. \ref{fig:ref_properties_2d} explains that $F$ has not reached an asymptote in the simulated domain. As a side note, Andrews et al. \cite{andr22} link the presence of a plateau in $F$ to beam detachment, which is not the case: Fig. \ref{fig:basic_ideal_properties} already showed that ion detachment is progressive and was evident already at $z=2R_0$.

Table \ref{Table:End_values_reordered3} compares the performance parameters of simulations B0 to B4, for TCA and OFW conditions. From B0 to B4 there is 
$\sim 13$\% increase in the ion mass flow, $\sim 19$\% of the beam power spent in $p_{bg}$-related ionization, and $\sim 10$\% decrease in the downstream ion velocity at the axis. Our results agree well with the experimental evidence of \cite{vial17,caru18,wach20a,desa23a}. Quantitatively, and  for the closest case of \cite{wach20a}, we have lower variations of the above variables (about a factor of 1.5 to 2). The difference can be due to the limited size of the simulation domain or  the choices made for certain parameters, such as $M_{i0}$, $T_n$, and  $\gamma_e$.   
A higher $p_{bg}$ implies that ions are hotter (up to 0.5 eV in simulation B4
--- as the single-fluid, Maxwellian temperature value), but their thermal energy continues to be marginal. As expected, the plume divergence increases with $p_{bg}$.  The fact that the  wall potential $|\phi_W|$ is independent of  $p_{bg}$ gives reliability to  OFW conditions.

Differences between applying TCA and OFW conditions are worth commenting on, since they are related to the influence of the electrical downstream conditions on the whole plasma response. Taking simulation B4 for illustration, MN performances are worse for  OFW conditions than for the TCA ones: the plume divergence is  10\% higher and more power is spent on ionization, but the final thrust is lower. Indeed, while $F_P$  increases  with $p_{bg}$ (up to  9\% from B0 to B4) when using the TCA conditions, it decreases slightly when using OFW conditions. All experimental evidence supports this last behavior,  which would reassert that OFW conditions are most physical and appropriate.

Finally,  Table \ref{Table:End_values_reordered3} also gathers the value  $\bar\gamma_e$  for the polytropic law \eqref{eq01} obtained from a linear  regression on the curves  $\log T_e$ versus $\log n_e$.   From B0 to B4,  $\bar\gamma_e$ goes from 1.2 to 1.28, well within the range of experimental fittings. The variation in the computed parameter is mainly observed in the low-pressure regime, from B0 up to 1 mPa, while minor changes are obtained between B2 and B4.

%%%%%%%%%%%%%%%%%%%%%%%%
\section{Further considerations}\label{sec:model_validation}
%%%%%%%%%%%%%%%%%%%%%%%

\subsection{On the size of the simulation domain}

In our previous simulations, the cylindrical simulation domain is  just 18 cm × 18 cm, much smaller than an actual testing chamber. 
An advantage of using the OFW condition over the TCA one is that the region between P and W —treated above as a Debye sheath—can also be interpreted as a downstream matching layer (DML)  between the simulation boundary P and the actual chamber wall, or the infinity  downstream in case of expansion into vacuum, as proposed in \cite{domi25a}. The main phenomenon expected in the DML is that the potential drop between P and W accommodates the in-plane electron current reaching W in order that the net electric  current of the plasma plume is zero \cite{ahed20a}. 

Due to the limited size of the simulation domain, it is  pertinent to verify that the results obtained for a size $L/R_0=10$  (including the potential at W) are robust against changes in the domain size. Figure \ref{fig:domain_analysis} compares results
for simulation B4 and a new simulation B4e  with an extended domain $L/R_0=15$ and the same BCs. 
The first row compares  $n_e$, $T_e$, and $\phi$ along the axis for both simulations. The agreement  is  total for the TCA case, when all conditions are imposed at the throat, and the numerical integration has no information about the domain size.  
The agreement is not total, but excellent, for the OFW case. 
The variables  in the second row  of Fig. \ref{fig:domain_analysis}  already contain information about the outer boundary, and the agreement between B4 and B4e remains excellent for both TCA and OFW. 
Table \ref{Table:End_values_reordered3} compares the performance  for B4 and B4e. As the domain is enlarged, the ratios $\dot{m}_{iP}/\dot{m}_{i0}$, $P_P/P_0$ and the thrust-related quantities are nearly identical between both simulations, with differences below 2\%. 
%Therefore, the simulation B4 can be considered robust.

Nonetheless, from $z/R_0=10$ to 15, residual thrust and power are still transferred from electrons to ions, and $\phi$ continues  decreasing. Still, $\phi_D-\phi_W\approx$ 20 V and $T_{eD}\approx $2 eV,  which means that the expansion is not fully complete, which also explains that the profiles of $F/F_0$ in Fig. \ref{fig:domain_analysis} have not reached an asymptote yet.  

{The physics of the region between P and W includes other phenomena beyond the development of the suitable potential drop. First, there is the progressive electron demagnetization, to be discussed in the next subsection. Then there is the magnetic field induced by the azimuthal plasma
currents, which opposes the applied magnetic field and whose relative importance increases downstream, therefore contributing to electron demagnetization.  Its spatial distribution could be computed from the Biot-Savart
law and added iteratively to the present model in the way proposed in \cite{meri16b}. Its relative strength is estimated with the
beta-parameter, $\beta=\mu_0 p_e/B^2$. In our simulation domain it turns out to be small, justifying neglecting the induced field: it is $\beta= 2.6\cdot10^{-3}$ at $(0,0)$ and $\beta=4.5$ - $6.5 \cdot10^{-2}$ at $(10R_0,0)$  for the simulations of Table \ref{Table:End_values_reordered3}.

%%%%%%%%%%%%%
\subsection{On the integration of the electron fluid model}
%%%%%%%%%%%%%

The validity of our integration scheme along magnetic lines for 
\eqref{eq:electroncontinuity_red} and \eqref{eq:electronenergy_red} requires
\begin{equation}
\bar\chi\ll 1,
\qquad
    u_{\perp e}=\bar\chi u_{\theta e}\ll u_{\parallel e},
 \qquad 
      \frac{(\gamma_e-1)}{2\gamma_e}\frac{m_eu_e^2}{T_e} 
    \approx \frac{m_eu_e^2}{10\,T_e}\ll 1.
    \label{eq03}
\end{equation}
Fig. \ref{fig:sketch} (b) plotted the Hall parameter $\chi\equiv \bar \chi^{-1}$ for simulation B4e. 
It is $\chi\sim 15800$  at (0,0), and $\chi\approx 275$ at $(15R_0,0)$. 
The two other conditions in Eq. \eqref{eq03} are plotted in  Fig. \ref{fig:inertial_terms} for simulation B4e, and TCA and OFW cases.
They are well satisfied within the main plasma beam for $B_0=400$ G.
Since $\bar\chi u_{\theta e}$ and $u_{\theta e}^2$ scale proportional to $1/B_0^2$ and $u_e\sim u_{\theta e}$, values on these plots scale  approximately as $1/B_0^2$. The maxima on these plots are at their top-right corner  and are  the result of  $u_{\theta e}$ increasing almost proportionally to $r$ \cite{ahed10f}; fortunately, it is a region of marginal interest. 
In terms of $\chi$, the simulations B4 and B4e are the most critical, since for simulation B0, $\chi\approx 6500$ at  $(15R_0,0)$.
Therefore, the validity of our integration scheme is well satisfied within our simulation domain for all cases. 

{For $\chi\leq 100$,  the set of electron equations \eqref{eq:electroncontinuity}-\eqref{eq:electronenergy} is better integrated with conventional finite-difference and finite-volume methods, preferably  in a magnetic field-aligned mesh, as Hall2De and HYPHEN do \cite{mark20a,zhou22a}. These methods are applicable also to the region $\chi<1$, as long as the drift-diffusive electron model continues to be valid. However, due to the large stiffness of the matrices involved, they tend to be nonconvergent numerically for very large $\chi$, that is, in the range of the numerical method proposed here. }

{A last and incidental observation is the following.} Beyond the very valuable experimental results, Wachs and Jorns \cite{wach20a} have proposed a \textit{global} model of  the plasma expansion on a MN, based on using only averaged values of plasma variables in the whole MN. However, models and experiments show the highly inhomogeneous, 2D character of the plasma expansion:  first, plasma densities and the ionization rate $R_{ion}(T_e)$ decrease by 2 or more orders of magnitude as the plasma expansion evolves;
second, the azimuthal current and the axial magnetic force are fully 2D, with local maxima at intermediate radii within the MN. With these features, we find it unaffordable to obtain approximate analytical expressions for the integrals on the right-hand sides of  Eqs. \eqref{eq:thrust_mass_P}-\eqref{eq:P_P}, which constitute the core of a global model. This type of model works reasonably well, for instance, in the discharge chamber of an ion thruster \cite{LIEB05}, where variations of plasma variables are rather moderate (say a factor of 2). 

\subsection{{On  momentum and thermal anomalous diffusion}}

In magnetized plume expansions, there are two phenomena, of very different nature, that can lead to 'anomalous' electron diffusion in an axisymmetric  fluid formulation.
First,  there is the cross-field diffusion due to nonlinear azimuthal instabilities.  Hepner et al. \cite{hepn20b}, Takahashi et al. \cite{taka22a}, and Maddaloni et al. \cite{madd25a} have  experimentally detected azimuthal oscillations in MN expansions, which could be a signature of this effect being present in MNs. 
{The phenomenon is well known to happen in Hall thrusters, but its kinetic or fluid nature is still under debate.
Second, there is the well-reported kinetic electron cooling in the MN plume, unrelated to instabilities, and due to the development of a highly non-Maxwellian electron velocity distribution function in the plume \cite{mart15a, ahed20a}. }

{These two phenomena can be included phenomenologically in an axisymmetric electron fluid model through fitted resistive terms in the momentum and energy equations. {Marks} et al. \cite{mark20a} claim the need for anomalous resistivity to reproduce the observed electron cooling, but they justify it on the instabilities of \cite{hepn20b}. Their approach consists first of substituting $\nu_e$ by an effective $\nu_e^*=\nu_e+\alpha_{ano}eB/m_e\gg\nu_e$, with $\alpha_{ano}$ a parameter (or variable) to be fitted. This defines 'anomalous' values for the mobility $\mu_e^*$ and the Hall parameter $\chi^*$ in Eq.\eqref{eq04}, with $\mu_e^*\ll \mu_e$ and $\chi^*\ll \chi$. Then, the cross-field electron current and the parallel heat flux behave, respectively and approximately as
\begin{equation}
    j_{\perp e}\propto \frac{\mu_e^*}{1+(\chi^*)^2}\propto \frac{1}{\chi^*}, 
    \qquad 
    q_{\parallel e} \propto \mu_e^* \nabla_\parallel T_e,
\end{equation}
implying an important enhancement of  the cross-field current and the parallel gradients of $T_e$.
The problem is that both phenomena are linked to a single parameter.}

{Instead,  Zhou et al. \cite{zhou22a}, using HYPHEN, suggested treating cross-field diffusion and parallel electron cooling separately. Their model yields \cite{ahed23a}
\begin{equation}
    j_{\perp e}\propto \frac{\mu_e}{1+\chi\chi^*}\propto \frac{1}{\chi^*}, 
    \qquad 
    q_{\parallel e} \propto \mu'_e\nabla_\parallel T_e,
\end{equation}
with $\mu'_e$ being tuned independently of $\alpha_{ano}$ (and $\mu'_e=\mu_e$ if no kinetic cooling is assumed).
Recently, this model has also been tested successfully by
Fernández et al. \cite{fern26a} for a cylindrical Hall thruster with a downstream MN. However,  they also reported the development of  a higher peak of $T_e$ inside the plasma source and larger electron parallel currents, these ones affecting the numerical convergence. }

{These issues, plus the previously cited  theoretical evidence on the convective behavior of the parallel energy flux in these MNs, have led us to propose and test the convective electron energy equation \eqref{eq:electronenergy} as a suitable empirical closure. Cross-field diffusion has not been included mainly to show that it is not strictly necessary  for plume cooling. The effect of adding it in plume divergence is near-equivalent to decreasing $B_0$ adequately.}

{Further kinetic studies should help to define a more consistent  closure of the electron fluid model. For instance, the kinetic investigations of paraxial MNs by Zhou in \cite{zhou21} (section 6.5) led him to propose a combined convective-conductive heat flux law, and he showed that the relative importance of convection increases as collisionality decreases. The PIC simulations of Andriulli et al. \cite{andr24} use an anomalous collisionality $\nu_e^*$ on each macro-electron, implying  that $\chi^*\leq $ 64 and not allowing us to distinguish the cooling due to $\nu_e^*$ from the one due to the partial depletion of the electron velocity distribution function.}

%%%%%%%%%%%%%%%%%%%%%%%%
\section{Conclusions}\label{sec:conclusions}
%%%%%%%%%%%%%%%%%%%%%%%

A stationary, axisymmetric three-fluid formulation
of the plasma quasineutral expansion in a magnetic nozzle at the high electron magnetization limit has been presented to analyze  in-plume ionization of the emitted plasma beam, background pressure effects, and the electrical connection between the MN throat and the outer boundary. 

It is shown that the presence of moderate collisions does not alter qualitatively the main plasma physics along the magnetic nozzle: the ambipolar electric field efficiently transfers  electron thermal energy to ion directed kinetic energy, while electrons remain strongly magnetized, ions progressively detach from magnetic lines due to their inertia, and the azimuthal electron current (combination of diamagnetic and $E\times B$ drifts) generates magnetic thrust on the thruster. 
But collisional processes, particularly ionization, leave their mark on the plasma response. First, they take energy out of the electrons, augmenting their cooling, thus reducing the potential fall along the nozzle and the ion downstream velocity. Simultaneously, they increase the ion mass flow and thus the thrust, and since new ions are born almost at rest, they are more prone to radial deflection and increase plume divergence. 
When ionization is due to background neutrals, instead of ejected ones, the effects are more distributed along the plume, as expected. 
{The simulations do not try to replicate a particular experiment, 
which would require having enough data to fix all the input parameters of the model.
Nonetheless, the trends related to thrust, plume divergence, and ion final velocity, with background pressure match well with experimental observations in the literature. } 

The proposed formulation of the electron fluid model differs from standard ones by postulating the dominance of electron energy convection versus diffusion. 
This has allowed us to reproduce the experimentally observed plume cooling without including any anomalous cross-field or parallel diffusion of momentum or energy in the model, and without imposing the downstream electron temperature.  The differences between anomalous electron cross-field  diffusion
and anomalous electron cooling,  the ways to model them empirically in an axisymmetric  fluid formulation, and the theoretical support of these models have been commented. A more detailed comparison of the convective and diffusive energy closures is left for future work.

The numerical integration scheme applied to the electron equations extends the  one in the collisionless limit of DIMAGNO. The three electron scalar variables (the electron flow, a thermalized potential, and an adiabaticity function)
that were constant along each magnetic line are now evolving along it due  to elastic and ionization collisions.   This integration scheme is estimated to be valid for Hall parameters higher than 100, roughly. For lower values of the Hall parameter, cross-field diffusion is not negligible, and standard finite-difference or finite-volume methods are more suitable. In any case, the suitability of the drift-diffusion electron fluid model as demagnetization decreases is not discussed in this work.

An important part of the full plasma model is the boundary conditions, mainly those related to electrons at the external boundaries of the reduced simulation domain. 
The imposition of  local current ambipolarity at the MN throat is computationally simple (and thus widely used), but it is 
poorly supported by physical evidence. Instead, the global floating wall condition at the external boundary, combined with a downstream matching layer,  captures more adequately the conditions matching  electrically the MN throat (i.e., the thruster exit)   with a large metallic vacuum chamber. Interestingly, the two types of conditions present opposite trends in the evolution of the magnetic thrust with the background  pressure, with only the OFW showing a mild decrease in thrust as $p_{bg}$ increases, in agreement with the experimental evidence.

{Further advances towards consistency of the electron fluid model require, apart from more experimental data to compare with,  full 2D axial-radial kinetic simulations to better characterize the electron cooling, possibly with convective-conductive heat flux laws, and azimuthal spectral extensions \cite{ripo26a} to characterize the cross-field transport in MNs. }

%----------
\section*{Acknowledgments}
%----------------------------------------------------------------------

This work has been carried out as part of the ZARATHUSTRA project, which has received funding from the European Research Council (ERC) under the European Union's Horizon 2020 research and innovation programme (grant agreement No 950466). Additional support came from R\&D project PID2023-150052OB-I00 (ADAPT) funded by MICIU/AEI/10.13039/501100011033 and by ERDF, EU.

E. Ahedo has been supported by 
the R\&D project PID2022-140035OB-I00 (HEEP) funded by MCIN/AEI/ 10.13039/501100011033 and by “ERDF A way of making Europe”.

{The Department of Aerospace Engineering at Universidad Carlos III de Madrid has been accredited as a Unit of Excellence “María de Maeztu”, a distinction awarded by the Spanish State Research Agency for the period 01/01/2026-31/12/2031.}

\appendix

%---------------------------------------
\section{Numerical Integration Scheme} \label{sec:nmerical_integration}
%----------------------------------------------------------------------

In DIMAGNO-DG, the set of partial differential equations for ions and neutrals is cast as a single system in conservative form:
\begin{align}
\frac{\pd \bm Q}{\pd t} + \nabla \cdot \mathcal{\bm F} = \mathcal{\bm R},
\label{eq:QFR}
\end{align}
where 
\begin{align*}
    \bm Q = \begin{bmatrix}{}
             n_i,&
             m_in_i
             \bm u_{i},&
             n_i(\mathcal{E}_i+e\phi),&
             n_n,&
             m_in_n \bm u_{n},&
             n_n\mathcal{E}_n
        \end{bmatrix}^\top
\end{align*}
is the vector of unknowns, and
%$\mathcal{F}(\bm Q)$ and $\mathcal{R}(\bm Q)$ are the flux tensor and the vector that contains the right-hand side forcing terms, respectively, 

\begin{align*}
\quad
        \textbf{$\mathcal{F}$} = \begin{bmatrix}
             m_i n_i u_{zi} & m_in_iu_{ri}\\
             m_i n_i u_{zi}^2 + p_e + p_i & m_i n_i u_{ri}u_{zi}\\
             m_i n_i u_{zi}u_{ri} & m_i n_i u_{ri}^2 + p_e + p_i\\
             m_i n_i u_{zi}u_{\theta i} & m_in_iu_{ri}u_{\theta i}\\
             (\mathcal{E}_i+e\phi + T_i)n_iu_{zi} & (\mathcal{E}_i+e\phi + T_i)n_iu_{ri}\\
             m_i n_n u_{zn} & m_in_nu_{rn}
             \\
             m_i n_n u_{zn}^2 + p_n & m_i n_n u_{rn}u_{zn} + p_n
             \\
             m_i n_n u_{zn}u_{rn} + p_n & m_i n_n u_{rn}^2 + p_n
             \\
             (\mathcal{E}_n + T_n)n_nu_{zn} & (\mathcal{E}_n + T_n)n_nu_{rn}  
        \end{bmatrix},
        \quad\textbf{$\mathcal{R}$} = \begin{bmatrix}
            S_{ion}\\
            en_i(u_{\theta e} - u_{\theta i})B_r  + R_{zi}\\
            en_i( u_{\theta i}-u_{\theta e} )B_z  + R_{ri}\\
            en_i(u_{zi} B_r - u_{ri}B_z)\\
            Q_{in}+e\phi S_{ion}\\
            -S_{ion}\\
            -R_{zi}\\
            -R_{ri}\\
            -Q_{in}\\ 
        \end{bmatrix}
\end{align*}
are, respectively, the flux tensor and the forcing vector.
The numerical approach is analogous to the one used in \cite{meri23a}, although the problem is now extended to include neutral species, ion and neutral temperatures from total energy conservation, as well as ionization and charge exchange collisions.
The spatial discretization and numerical integration of the problem follow a Discontinuous Galerkin (DG) scheme of order $p = 0$, with the domain of interest, $\Omega$, discretized on a triangular mesh, $\mathcal{T}_h$, of 80,000 elements $\kappa$ of facet side length 1.3 mm. To formulate the problem in DG, the weak formulation is constructed by multiplying Eq. \eqref{eq:QFR} by the arbitrary smooth vectorial function $\textbf{v}$ and integrating by parts over $\kappa$ in the mesh. Then, all elements $\kappa$ in $\mathcal{T}_h$ are added together, 
\begin{align}
    \sum_{\kappa\in\mathcal{T}_h}\bigg[\int_\kappa \textbf{v}_h\cdot \frac{\pd \bm Q_h}{\pd t}d\textbf{x} - \int_\kappa \mathcal{F}:\nabla\textbf{v}_hd\mathbf{x}\notag
    \\
    +\int_{\partial \kappa} (\textbf{v}_h^+-\textbf{v}_h^-)\cdot\mathcal{H}_{LF}\cdot \mathbf{1}_ndS = \int_\kappa \textbf{v}_h\cdot\mathcal{R}d\mathbf{x}\bigg],\label{eq:weak_form}
\end{align}
where superindices '$+$' and '$-$' indicate internal and external facets of the cell, $\mathbf{1}_n$ is the unitary normal vector of the facet pointing outward ('$+$'), and $\mathcal{H_{LF}}$ is a numerical flux function chosen, in the present study, to be the local Lax-Friedrichs flux, given by
\begin{equation}
    \mathcal{H_{LF}} = \frac{1}{2}[\mathcal{F}(\bm Q^+)-\mathcal{F}(\bm Q^-)+\alpha (\bm Q^+-\bm Q^-)],
\end{equation}
with $\alpha$ computed as the maximum of all eigenvalues of the normal flux Jacobian ($\nabla\mathcal{F}\cdot \bm 1_n$) evaluated on each side of the facet \cite{meri23a}. The problem is then evolved in time with a strong stability preserving Runge-Kutta (SSPRK) method \cite{shu88}, until a steady state is reached. The convergence criterion of the solution for a steady state is given by
\begin{align}\label{eq:convergence}
    \sum_{\forall\text{ nodes}} \sum_{\forall i} \frac{|\bm q_{i,n+1}- \bm q_{i,n}|}{t_{n+1}-t_n}<10^{-8},
\end{align}
with $\bm q_{in}$ being the non-dimensional components of $\bm Q$, and the subindex $n$ the temporal instant.

In order to initialize the simulation, a map of $A$ and $\Phi$ must be fixed before solving the equations for the heavy species. To this end, we ignore the collisional effects of ionization and charge-exchange on equations \eqref{eq:G_integration}-\eqref{eq:Phi_integration}, which return constant functions along magnetic streamlines. We analytically calculate the value of the magnetic streamfunction $\psi$ and find that the values of $A$, $\Phi$, and $u_{\theta e}$ can be fixed everywhere by interpolation from the upstream conditions. Note that, by doing so, the solution for $G = nu_{\parallel e}/B$ is initially decoupled from the plasma model.

At $t = 0$, the simulation is initialized with a low-density background (i.e., $n_i = 10^{-6}n_{e0}$) and, after a sufficient number of integration steps, convergence is reached. Once the first steady state is achieved,  $G$, $A$, and $\Phi$ are updated by performing one-dimensional integration along magnetic lines with Eqs. \eqref{eq:G_integration}, \eqref{eq:C_integration}, and \eqref{eq:Phi_integration}, respectively. {The} simulation is then re-initialized with the previous steady-state solution and updated values for electrostatic potential and collisional effects, expecting subsequent convergence.

%%%%%%%%%%%%%%%%%%%%%%%%%
\section{Computation of performance variables}
\label{AppB}
%%%%%%%%%%%%%%%%%%%%%%%%%%

Starting with plasma power, adding the energy equations of neutrals, ions, and electrons, the plasma total energy flux $\bm P''$ 
satisfies

\begin{equation}
    \nabla\cdot \bm P''=-S_{ion}\mathcal{E}_{inel} - \bm j\cdot \nabla \phi,
    \qquad
    \bm P''(z,r)=\sum _{\alpha=n,i,e} \bm P''_\alpha,  
    \qquad
        \bm P''_\alpha =(\mathcal{E}_\alpha+ T_\alpha)n_\alpha\bm u_\alpha.
\end{equation}
Writing $\bm j\cdot \nabla \phi=\nabla\cdot\bm j\phi$ for $\nabla\cdot\bm j=0$,  integrating the above differential equation over the whole simulation domain $\Omega$ and taking into account the BCs on $\phi$ and $\bm j$, the power stationary balance is 
\begin{equation*}
    P_P-P_0=-\int_{\Omega} \dd\Omega\ S_{ion}\mathcal{E}_{inel},
\end{equation*}
with $\int_{\Omega} \dd\Omega \equiv 2\pi \int_{0}^{L} dz \int_{0}^{L} dr r $, $P = \sum _{\alpha = n,i,e}P_{\alpha}$ and
\begin{align}
&
    P_{\alpha P}=2\pi \Bigg[
     \int_{0}^{L} \dd r \, r\, P''_{z \alpha}(L,r)
     + L \int_{0}^{L} \dd z \, P''_{r \alpha}(z,L)
     - \int_{2R_0}^{L} \dd r \, r\, P''_{z\alpha}(0,r)
\Bigg],
\\
&
    P_{\alpha 0}=2\pi \int_{0}^{2R_0} \dd r \, r\, P''_{z\alpha}(0,r),
\end{align}
which are surface integrals of $\bm P''$ over the throat and the outer boundary

Proceeding in the same way with the ion continuity equation  \eqref{eq:ioncontinuity}, the  stationary balance of the ion mass flow satisfies
\begin{equation*}
    \dot m_{iP}-\dot m_{i0}=m_i\int_{\Omega} \dd\Omega\ S_{ion},
\end{equation*}
and the expressions for $\dot m_{iP}$ and $\dot m_{i0}$ are straightforward. Similar balances exist for other mass or electric flows.

In steady-state and vacuum, the plasma momentum flux $\bar{\bar{\bm M}}$  satisfies 
\begin{equation}
    \nabla \cdot \bar{\bar{\bm M}}=\bm j\times \bm B,
    \qquad \bar{\bar{\bm M}}= \sum_{\alpha=n,i,e}( n_\alpha \bm u_\alpha\bm u_\alpha +p_\alpha \bar{\bar{\bm I}}),
\end{equation}
Taking the $z$ component of this equation and integrating it over the cylindrical simulation domain $\Omega$,  we obtain $\int_V dV \ (\bm 1_z \cdot \nabla \cdot \bar{\bar{\bm M}})
    =\int_V dV  ( \bm 1_z \cdot\bm j\times \bm B)$ , that is 
\begin{equation*}
    F_P-F_0=\int_{\Omega} d\Omega\ (-j_\theta B_r),
\end{equation*}
with $F = \sum _{\alpha = n,i,e}F_{\alpha}$,
\begin{align}
&
    F_{\alpha P}=2\pi \Bigg[
\int_{0}^{L} dr \, r\, M_{zz\alpha}(L,r)
+L \int_{0}^{L} dz \, M_{zr\alpha}(z,L)
- \int_{2R_0}^{L} dr \, r\, M_{zz\alpha}(0,r)
\Bigg],
\label{eq:F_P}
\\
&
    F_{\alpha 0}=2\pi \int_{0}^{2R_0} dr \, r\, M_{zz\alpha}(0,r),
    \label{eq:F_0}
\end{align}
and
\begin{equation}
M_{zr\alpha}
=
m_\alpha n_\alpha \, u_{z\alpha} \, u_{r\alpha},
\qquad
M_{zz\alpha}
=
m_\alpha n_\alpha \, u_{z\alpha}^{2}
+ p_\alpha.
\end{equation}

%----------------------------------------------------------------------
\bibliography{bibtex/others,bibtex/ep2}  
%----------------------------------------------------------------------

\newpage
%----------------------------------------------------------------------

\renewcommand{\arraystretch}{1.3}

\begin{table}[h!]
\centering
\caption{Nominal simulation N. Throat boundary conditions for each plasma species and selected plasma parameters. Xenon propellant mass is used.
}
\label{Table:Initial_conditions}

\begin{minipage}{0.56\textwidth}
\centering
\begin{ruledtabular}
\begin{tabular}{lccc}
$\alpha$ & $n$ & $i$ & $e$ \\
\noalign{\vskip 2pt\hrule height 1pt\vskip 2pt}
$\dot{m}_{\alpha0}$ [$\mu$g/s] & 60 & 100 & -- \\
$\bar{n}_{\alpha0}$ [$10^{17}$ m$^{-3}$] & 7.7 & 3 & 3 \\
\noalign{\vskip 2pt\hrule height 0.1pt\vskip 2pt}
$T_{\alpha0}$ [eV] & 0.1 & 0.1 & 10 \\
$c_{\alpha0}$ [km/s] & 0.35 & 3 & -- \\
$M_{\alpha0}$ & 1 & 0.5 & -- \\
$u_{z\alpha0}$ [km/s] & 0.35 & 1.49 & 1.49 \\
\noalign{\vskip 2pt\hrule height 0.1pt\vskip 2pt}
$P_{\alpha0}$ [W] & 0.015 & 0.130 & 4.39 \\
$F_{\alpha0}$ [$\mu$N] & 33 & 154 & 490 \\
\end{tabular}
\end{ruledtabular}
\end{minipage}
\hfill
\begin{minipage}{0.40\textwidth}
\centering
\begin{ruledtabular}
\begin{tabular}{lc}
Model and plasma parameters & \\
\noalign{\vskip 2pt\hrule height 1pt\vskip 2pt}
$R_0$ [cm] & 1.8 \\
$L/R_0$    & 10 \\
$B_0$ [G] & 400 \\
$\gamma_e$ & 1.2 \\
$\epsilon$ & $10^{-3}$ \\
$P_0$ [W] & 4.53 \\
$F_0$ [$\mu$N] & 678 \\
$\dot m_{i0}/\dot m$ & 0.625 \\
$F_0^2/(2\dot{m}P_0)$ & 0.32 \\
\end{tabular}
\end{ruledtabular}
\end{minipage}
\end{table}

\begin{table*}[h!]
\caption{Global performance parameters for the nominal simulation (N) and the
background-pressure simulations B0--B4 (0 to 4~mPa). Simulation B0m corresponds
to B0 with $M_{i0}=1$; B4e corresponds to B4 with an extended domain $L/R_0 = 15$.
Subscript $D$ denotes the outlet point $(z,r) = (L,0)$. Except for simulation N
(see Table~\ref{Table:Initial_conditions}), $F_0 = F_{i0} + F_{e0} = 645\ \mu\text{N}$.
Results are reported for the Throat local Current-Ambipolarity condition (TCA)
and the Outer-boundary Floating Wall condition (OFW).}
\label{Table:End_values_reordered3}
 
\begin{ruledtabular}
\begin{tabular}{l cc@{\hspace{20pt}} c@{\hspace{20pt}} cc@{\hspace{20pt}} cc@{\hspace{20pt}} cc@{\hspace{20pt}} c}
& \multicolumn{2}{c}{\hspace{-20pt}\textbf{N}}
& \hspace{0pt}\textbf{B0}
& \multicolumn{2}{c}{\hspace{-20pt}\textbf{B2}}
& \multicolumn{2}{c}{\hspace{-20pt}\textbf{B4}}
& \multicolumn{2}{c}{\hspace{-20pt}\textbf{B4e}}
& \hspace{0pt}\textbf{B0m} \\
& TCA & OFW & --
& TCA & OFW
& TCA & OFW
& TCA & OFW
& -- \\
\noalign{\vskip 2pt\hrule height 1pt\vskip 2pt}
$\dot{m}_{iP}/\dot{m}_{i0}$
  & 1.18  & 1.21  & 1     & 1.09  & 1.08  & 1.16  & 1.12  & 1.16  & 1.12  & 1     \\
$P_P/P_0$
  & 0.885 & 0.872 & 1     & 0.905 & 0.903 & 0.83  & 0.80  & 0.82  & 0.79  & 1     \\
$F_P/F_0$
  & 1.45  & 1.42  & 1.41  & 1.42 & 1.39  & 1.44  & 1.39  & 1.46  & 1.40  & 1.51  \\
\noalign{\vskip 2pt\hrule height 0.1pt\vskip 2pt}
$P_{iP}$ [W]
  & 3.59  & 3.3  & 3.9  & 3.47  & 3.43  & 2.96  & 3.46  & 3.12  & 3.56  & 4.64  \\
$P_{eP}$ [W]
  & 0.412  & 0.644  & 0.62  & 0.742  & 1.03  & 1.09  & 1.28  & 0.973  & 1.14  & 0.212  \\[2pt]
\noalign{\vskip 2pt\hrule height 0.1pt\vskip 2pt}
$F_{iP}$ [$\mu$N]
  & 904  & 884  & 869  & 844  & 834  & 833  & 825  & 847  & 833  & 784 \\
$F_{eP}$ [$\mu$N]
  & 46  & 48  & 41.6    & 74.4  & 63.4  & 96  & 71.1  & 94.2  & 67.6  & 41.3  \\[2pt]
\noalign{\vskip 2pt\hrule height 0.1pt\vskip 2pt}
$u_{ziD}$ [km/s]
  & 6.82  & 6.93  & 7.91  & 7.25  & 7.39  & 6.93  & 7.02  & 7.06  & 7.24  & 8.08  \\
$T_{iD}$ [eV]
  & 0.40  & 0.30  & 0.025 & 0.44  & 0.40  & 0.56  & 0.48  & 0.56  & 0.49  & 0.02  \\
$T_{eD}$ [eV]
  & 2.31  & 2.42  & 3.27  & 2.45  & 2.53  & 1.97  & 2.06  & 1.90  & 1.86  & 3.56  \\
$-\phi_D$ [V]
  & 30.9  & 32.0  & 40.6  & 35.9  & 37.3  & 32.6  & 33.6  & 33.5  & 34.8  & 38.6  \\
$-\phi_W$ [V]
  & --    & 53.6  & 53.7  & --    & 53.7  & --    & 53.7  & --    & 53.6  & 54.8  \\
$\alpha_{\mathrm{div}}$ [deg]
  & 34.3  & 35.0  & 33.7  & 35.5  & 37.5  & 36.6  & 40.6  & 34.9  & 43.7  & 34.0  \\
$\bar{\gamma}_e$
  & 1.22  & 1.22  & 1.20  & 1.27  & 1.27  & 1.28  & 1.28  & 1.29  & 1.28  & 1.20  \\
\end{tabular}
\end{ruledtabular}
\end{table*}

\begin{figure}[H]
\centering
\includegraphics[width=0.9\columnwidth]{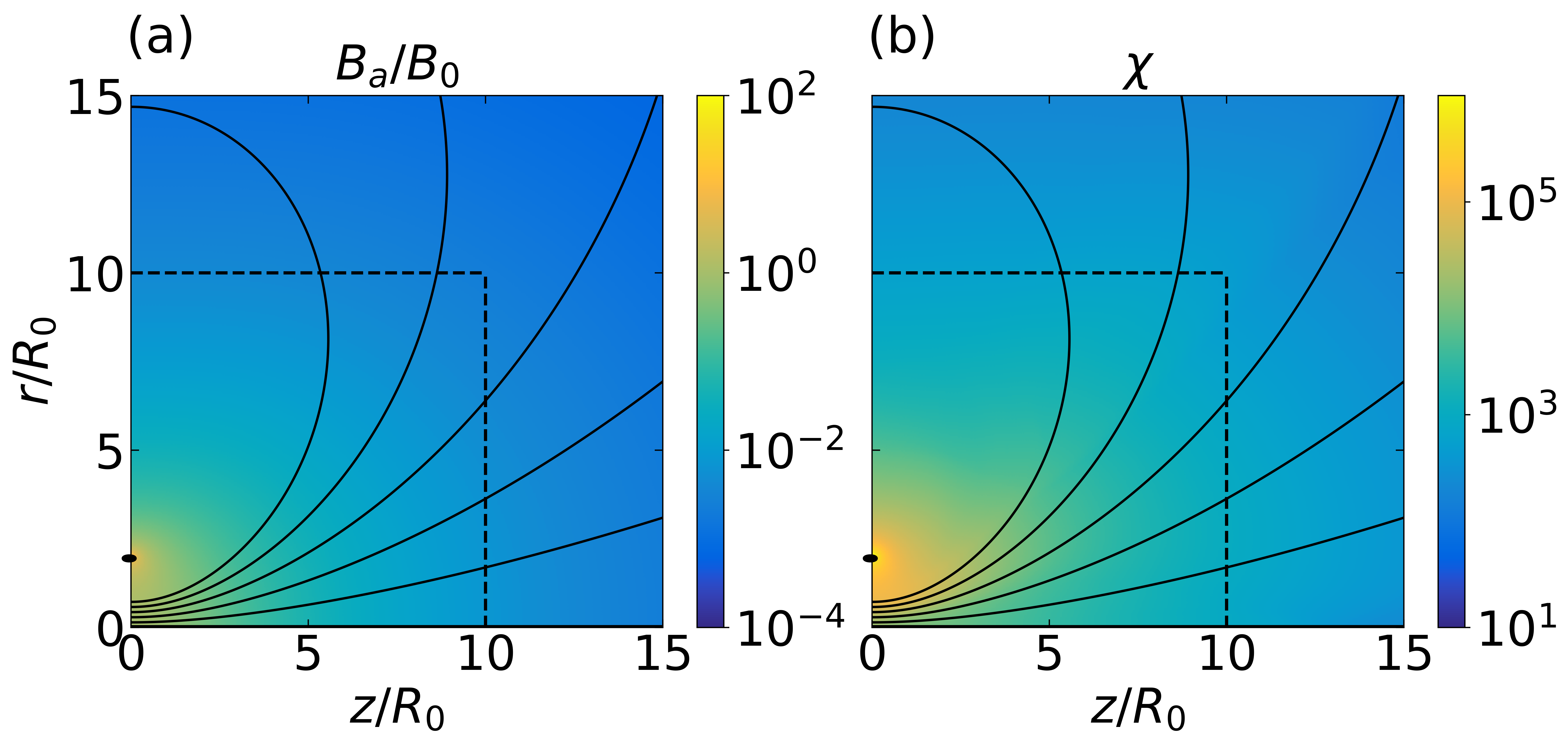}
\caption{Normalized applied magnetic field (a) $\bm B_a/B_0$ and Hall parameter (b) $\chi=eB/m_e\nu_{e}$ for simulations B4 (small square) and B4e (large square) in Table \ref{Table:End_values_reordered3}.  Black lines depict magnetic streamlines up to $r = 0.722R_0$, where 95\% of plasma is inputted. Black dot represents the single current loop at $(z,r) = (0,2R_0)$.}
\label{fig:sketch}
\end{figure}

\begin{figure}[H]
\centering
\includegraphics[width=1\columnwidth]{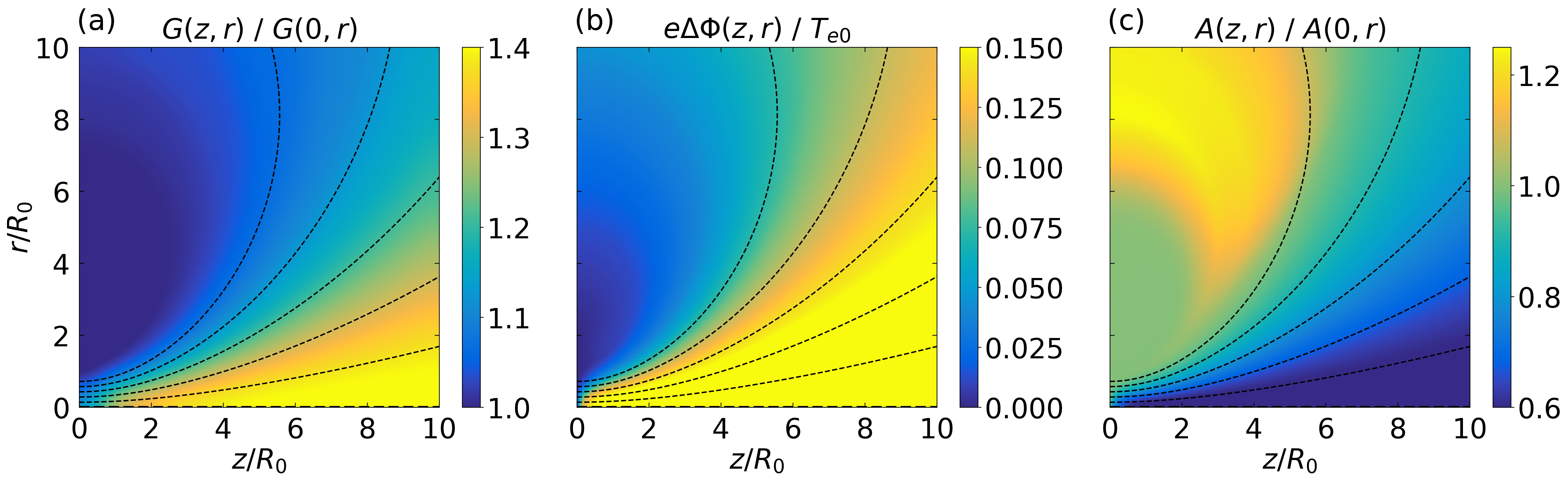}
\caption{Nominal simulation N with TCA condition. Maps for the electron functions (a) $G(z,r)/G(0,r)$, (b) $e\Delta \Phi(z,r)/T_{e0}$,  and  (c) $A(z,r)/A(0,r)$, with  $\Delta\Phi(z,r)=\Phi(z,r)-\Phi(0,r)$.
}
\label{fig:electronic_properties_2d}
\end{figure}

\begin{figure}[H]
\centering
\includegraphics[width=1\columnwidth]{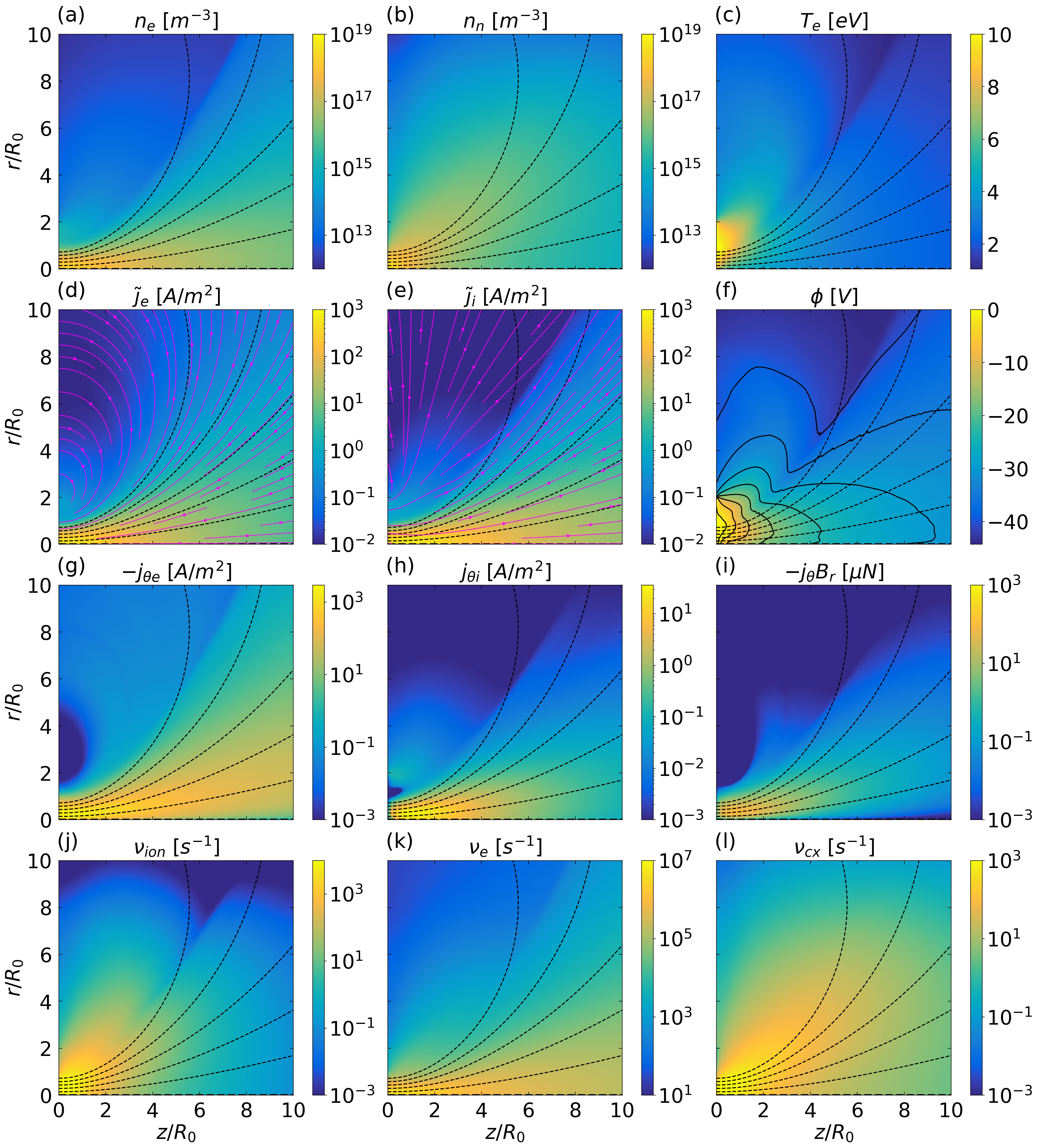}
\caption{Nominal simulation N with TCA condition. Maps of ion-electron and neutral densities (a) $n_e$ and (b) $n_n$; electron temperature (c) $T_e$; in-plane electron and ion currents (d) $\tilde{j}_e$ and (e) $\tilde{j}_i$; electrostatic potential (f) $\phi$; azimuthal electron and ion currents (g) $-j_{\theta e}$ and (h) $j_{\theta i}$; axial magnetic force (i) $-j_\theta B_r$; and collision frequencies (j) $\nu_{ion}$, (k) $\nu_e$, and (l) $\nu_{cx}$. 
Dashed lines are magnetic lines up to $r = 0.722R_0$. The map of $\phi$ includes contour lines separated by 5\ V.  The maps of $\bm{\tilde\jmath}_i$ and $\bm{\tilde\jmath}_e$ include both magnitude and streamlines.
}
\label{fig:ref_properties_2d}
\end{figure}

\begin{figure}[H]
    \centering
    \includegraphics[width=0.75\linewidth]{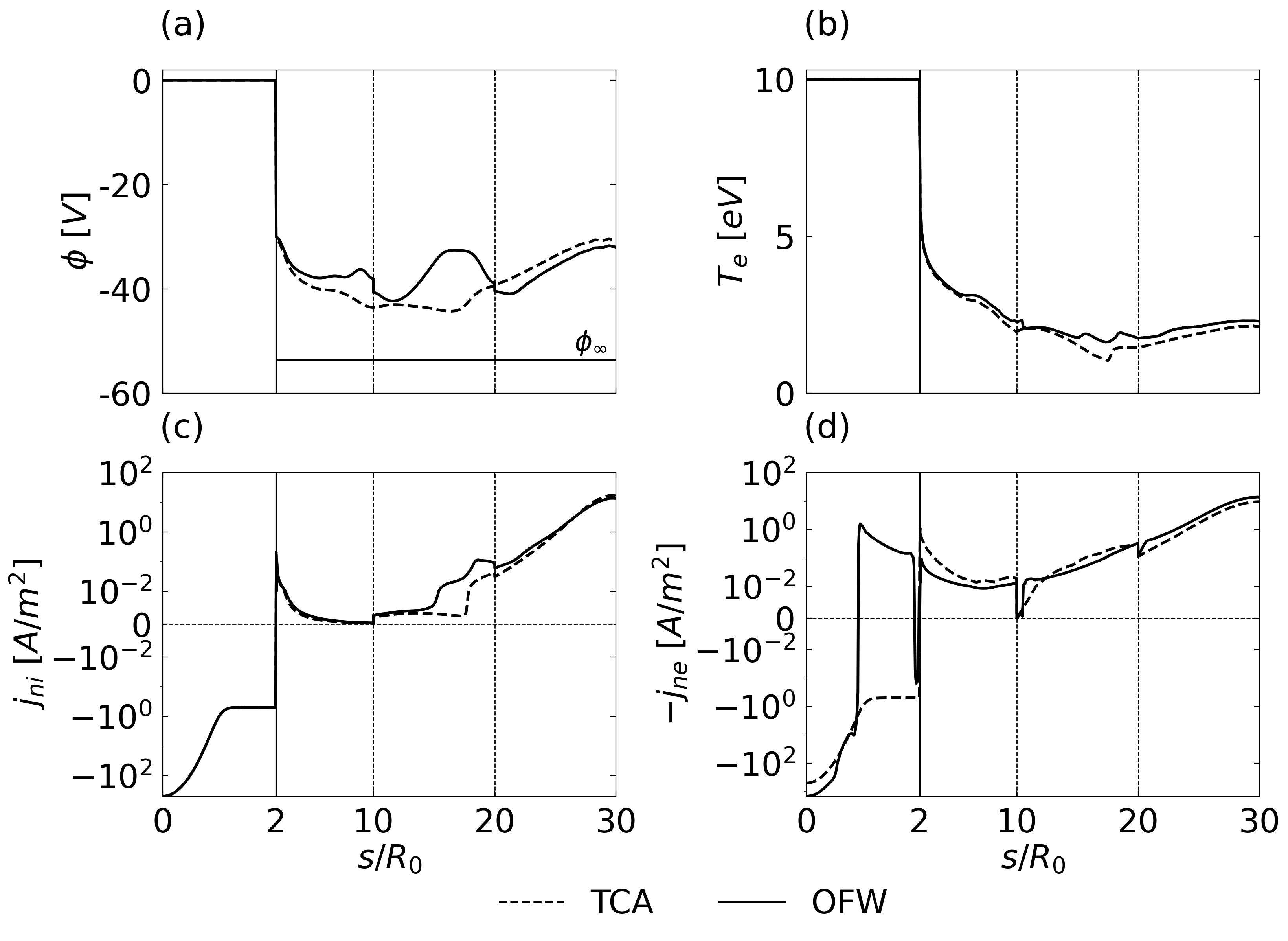}
    \caption{Nominal simulation N with TCA (dashed lines) and OFW (solid) conditions. Plasma variables at the domain boundaries (except for the axis): (a) $\phi$, (b) $T_e$, (c) $j_{ni}$, and (d) $-j_{ne}$. Arc length $s$ starts at $(z,r)=(0,0)$, follows the boundary and ends at $(z,r)=(L,0)$. 
     Note that the interval $0\leq s \leq 2$ in the horizontal axis, corresponding to the throat,  has been expanded for visibility; the vertical axis for $j_{ni}$ and $j_{ne}$ include both positive and negative values; and the sharp minima of $j_{ne}$ in the corners is likely of numerical origin.
    }
    \label{fig:boundaries_1d}
\end{figure}

\begin{figure}[H]
\centering
\includegraphics[width=1\columnwidth]{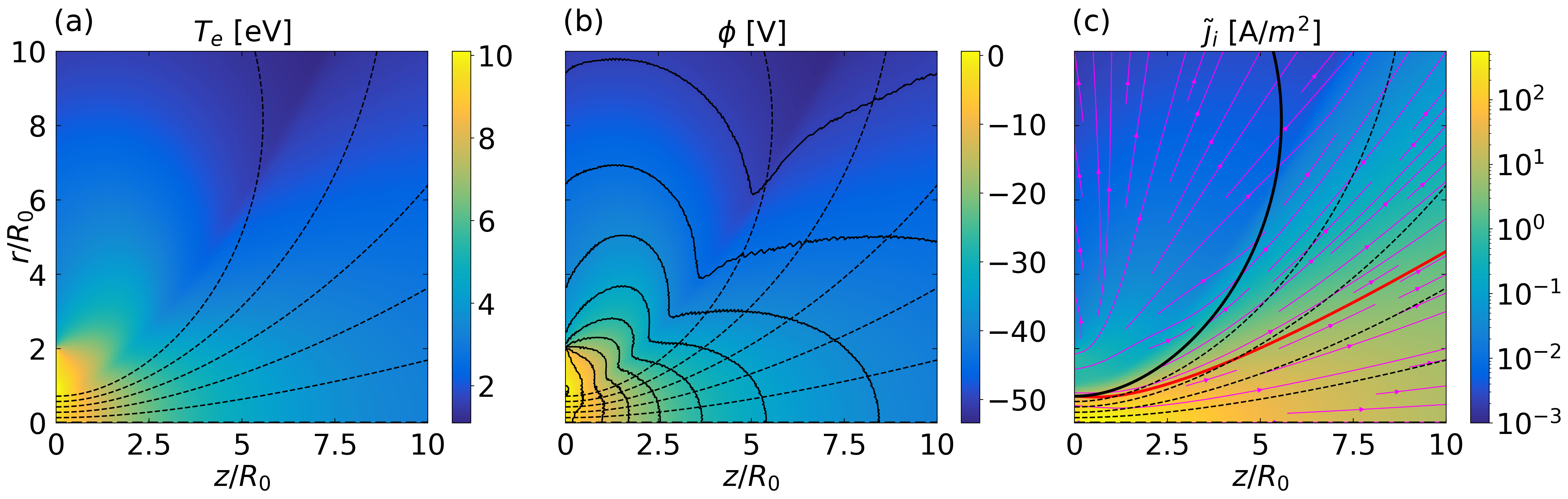}
\caption{Collisionless simulation B0.  Maps of (a) $T_e$, (b) $\phi$, and (c) $\tilde{j}_i$, to be compared  with the same ones for simulation N in Fig. \ref{fig:ref_properties_2d} In the last plot the red and black lines correspond to 95\% of the ion and (magnetized) electron flows, respectively. 
}
\label{fig:basic_ideal_properties}
\end{figure}

\begin{figure}[H]
    \centering
    \includegraphics[width=1\linewidth]{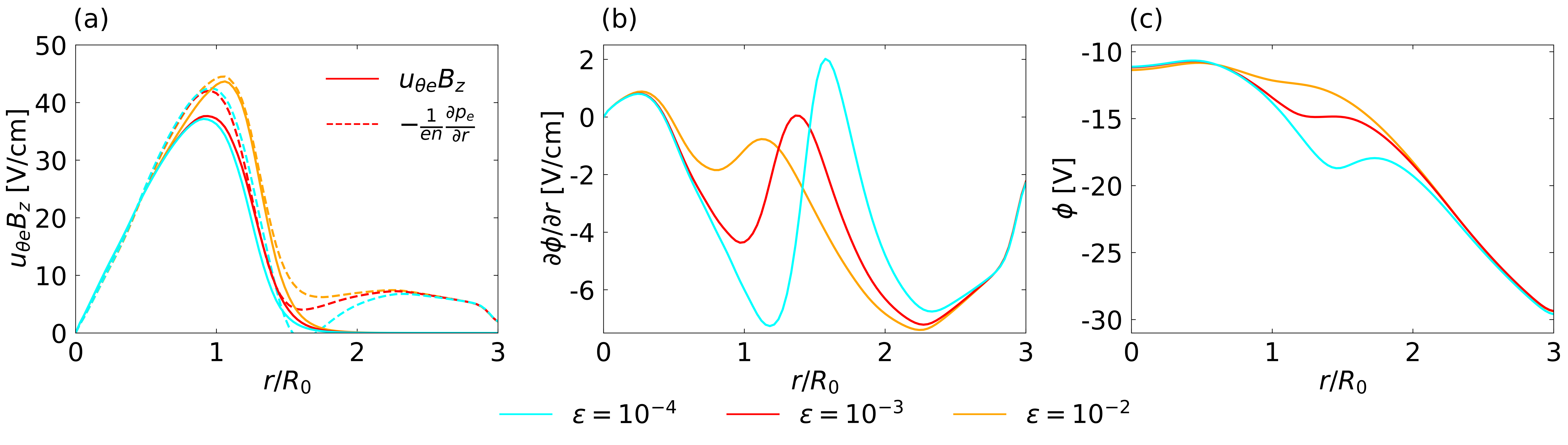}
    \caption{Collisionless simulation B0 with TCA condition. 
    Electron radial balance at $z=R_0$ for three levels of near-vacuum density $\epsilon n_{e0}$ in the region around the magnetic coil: (a) magnetic ($u_{\theta e}B_z$, solid lines) and pressure ($-(1/en)\partial p_e/\partial r$, dashed lines) terms, (b) radial electric field $\partial\phi/\partial r$, and (c) electric potential $\phi$.
    }
    \label{fig:phi_study}
\end{figure}

\begin{figure}[H]
\centering
\includegraphics[width=1\columnwidth]{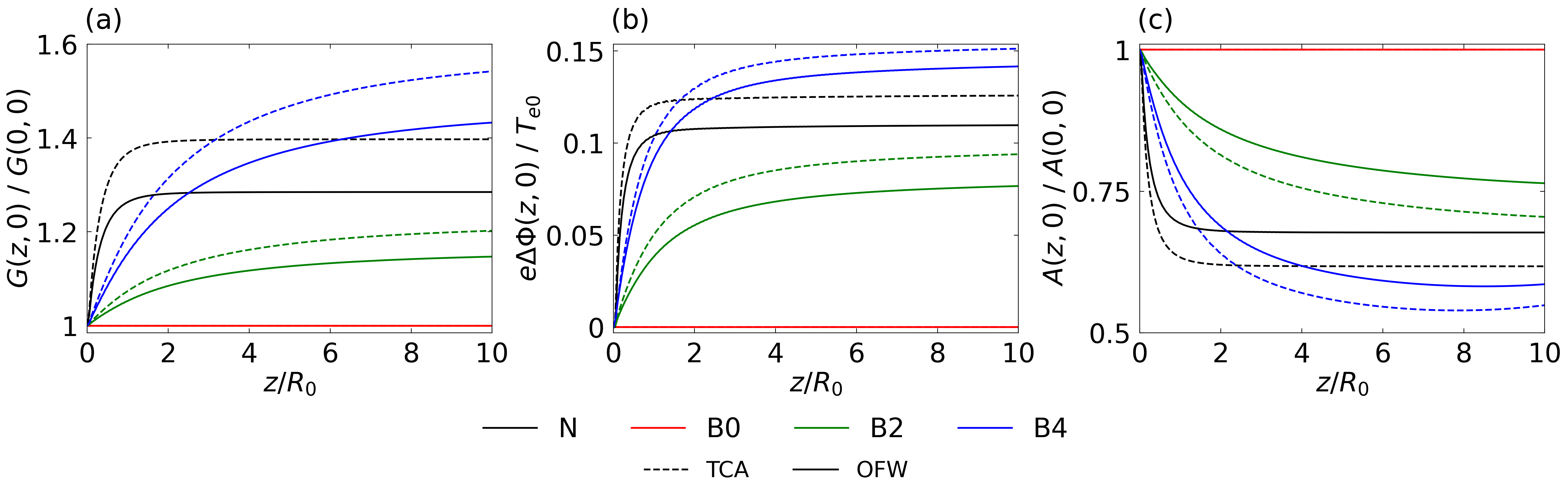}
\caption{Comparison of different simulations for TCA (dashed lines) and OFW (solid lines) conditions. 1D axial profiles at the axis of normalized electron functions (a) $G(z,r)/G(0,r)$, (b) $e\Delta \Phi(z,r)/T_{e0}$,  and  (c) $A(z,r)/A(0,r)$, with  $\Delta\Phi(z,r)=\Phi(z,r)-\Phi(0,r)$.}
\label{fig:electronic_properties_1d}
\end{figure}

\begin{figure}[H]
\centering
\includegraphics[width=1\columnwidth]{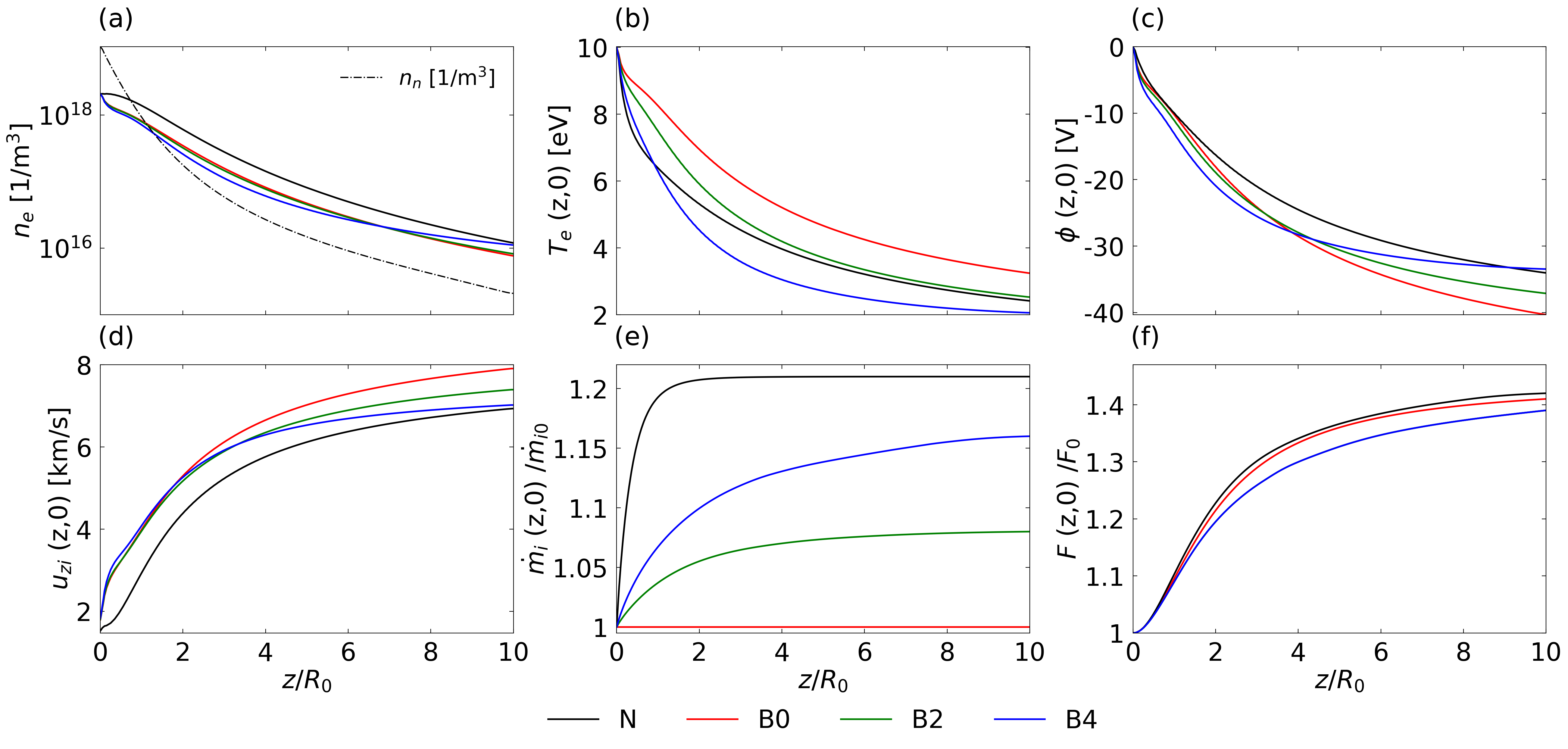}
\caption{Comparison of different simulations with the OFW condition.  
1D axial profiles at $r=0$ of (a) $n_e$, $n_n$ (dashed), (b) $T_e$, (c) $\phi$, and (d) $u_{zi}$, and 1D profiles of ion mass flow (e) and thrust (f), computed at $z=$const sections from integration over $0<r<L$.
}
\label{fig:properties_1d}
\end{figure}

\begin{figure}[H]
\centering
\includegraphics[width=1\columnwidth]{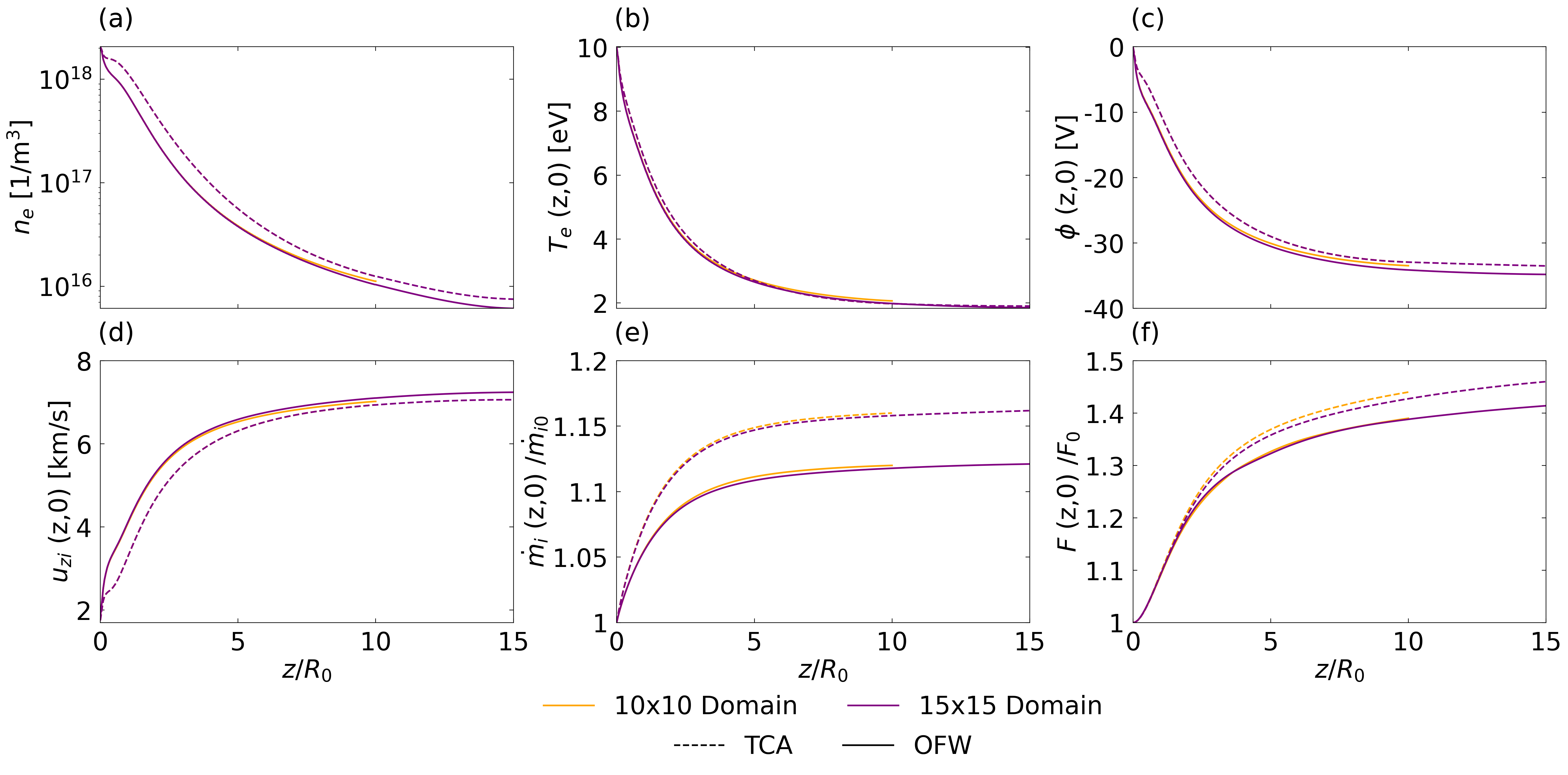}
\caption{Effect of the domain size for simulations B4 and B4e  with the TCA and OFW condition. Panels (a)--(f) show the same variables as in Fig. \ref{fig:properties_1d}. 
}
\label{fig:domain_analysis}
\end{figure}
\begin{figure}[H]
\centering
\includegraphics[width=0.7\columnwidth]{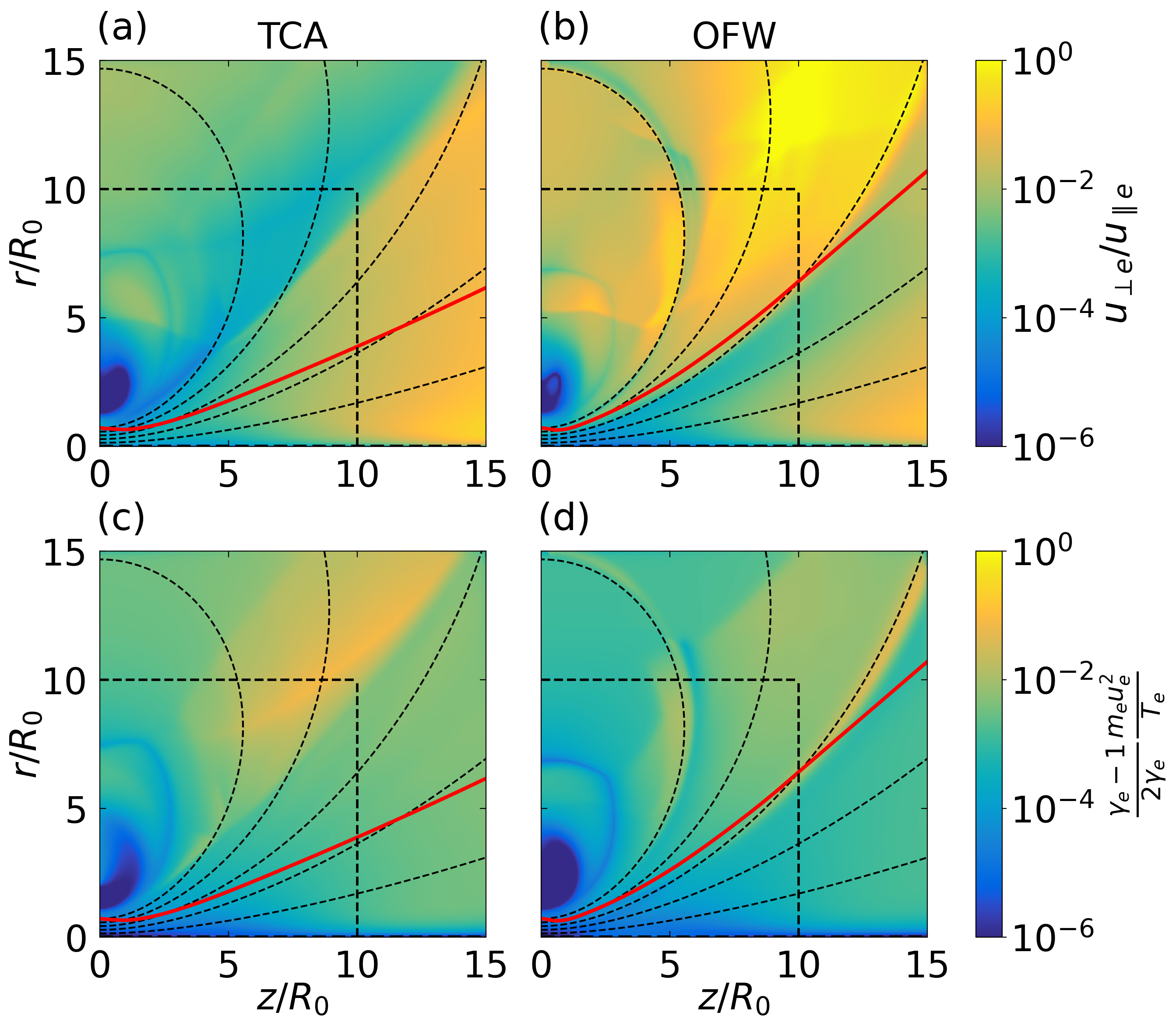}
\caption{Simulation B4 with TCA and OFW conditions. Ratios on electron variables for the verification of the magnetized electron model: $u_{\perp e}/u_{\parallel e}$ for (a) TCA and (b) OFW, and $(\gamma_e-1)m_e u_e^2/(2\gamma_e T_e)$ for (c) TCA and (d) OFW. The red lines correspond to the streamtube containing 95\% of the ion mass flow
{and evidence the larger plume divergence when applying the OFW conditions.}
}
\label{fig:inertial_terms}
\end{figure}

%----------------------------------------------------------------------
\end{document}